\pdfoutput=1

\documentclass[12pt]{article}

\usepackage{amssymb}
\usepackage{bm}
\usepackage{enumitem}
\usepackage{graphicx}
\usepackage{fullpage}
\usepackage{setspace}
\usepackage{hyperref}
\usepackage[font=small]{caption}
\usepackage{longtable}  % for tables across multiple pages in SI.
\usepackage{multirow}  % for tables with merged cells in SI.
\usepackage{verbatim}
\usepackage{totcount}
\usepackage{mathptmx}
\usepackage{xcolor}

\usepackage{times}
\usepackage{cite}

\newenvironment{sciabstract}{%
\begin{quote} \bf}
{\end{quote}}

 \newcommand{\beginsupplement}{%
        \setcounter{table}{0}
        \renewcommand{\thetable}{S\arabic{table}}%
        \setcounter{figure}{0}
        \renewcommand{\thefigure}{S\arabic{figure}}%
         \setcounter{section}{0}
        \renewcommand{\thesection}{S\arabic{section}}
         \setcounter{equation}{0}
         \renewcommand{\theequation}{S\arabic{equation}}
     }

\newcommand{\figref}[2]{Fig.\ \ref{#1}\uppercase{\textbf{#2}}}

\usepackage{xspace}
\newcommand{\NumCountries}{83\xspace}

\newcommand{\NumDisciplines}{13\xspace}
\newcommand{\YearStart}{1955\xspace}
\newcommand{\YearEnd}{2010\xspace}
\newcommand{\Population}{1,523,002\xspace}
\newcommand{\PopulationApprox}{1.5 million\xspace}
\newcommand{\PopulationMale}{1,110,194\xspace}
\newcommand{\PopulationFemale}{412,808\xspace}
\newcommand{\PopulationRaw}{7,863,861\xspace}
\newcommand{\PopulationRawMale}{2,146,926\xspace}
\newcommand{\PopulationRawFemale}{856,889\xspace}
\newcommand{\FemalePercentage}{27\%\xspace}
\newcommand{\CountryList}{Algeria, Argentina, Armenia, Australia, Austria, Bangladesh, Belarus, Belgium, Bolivia, Bulgaria, Cameroon, Canada, Chile, Colombia, Costa Rica, Croatia, Cuba, Cyprus, Czech Republic, Denmark, Ecuador, Egypt, Estonia, Finland, France, Gabon, Germany, Greece, Hungary, Iceland, India, Indonesia, Iran, Ireland, Israel, Italy, Jamaica, Jordan, Kazakhstan, Kenya, Kuwait, Latvia, Lebanon, Lithuania, Luxembourg, Macedonia, Madagascar, Mexico, Morocco, Netherlands, New Zealand, Nigeria, Norway, Pakistan, Peru, Philippines, Poland, Portugal, Qatar, Romania, Russia, Saudi Arabia, Senegal, Serbia, Slovakia, Slovenia, South Africa, Spain, Sri Lanka, Sweden, Switzerland, Tanzania, Thailand, Tunisia, Turkey, Uganda, Ukraine, United Arab Emirates, United Kingdom, United States, Uruguay, Uzbekistan, Venezuela\xspace}
\newcommand{\FemalePercentageEarly}{12\%\xspace}
\newcommand{\FemalePercentageRecent}{35\%\xspace}

\newcommand{\FemalePercentagePsychology}{33\%\xspace}

\newcommand{\FemalePercentageMathematics}{15\%\xspace}

\newcommand{\FemalePercentageGermany}{28\%\xspace}

\newcommand{\FemalePercentageRussia}{50\%\xspace}

\newcommand{\PubMale}{13.2\xspace}
\newcommand{\PubFemale}{9.6\xspace}
\newcommand{\PubGap}{27\%\xspace}
\newcommand{\PubGapTopAuthor}{37\%\xspace}

\newcommand{\PubGapEarly}{10\%\xspace}
\newcommand{\PubGapRecent}{35\%\xspace}
\newcommand{\PubGapChemistry}{35.1\%\xspace}

\newcommand{\PubGapAppliedphysics}{7.8\%\xspace}

\newcommand{\PubGapCountryPearson}{0.56\xspace}
\newcommand{\PubGapDisciplinePearson}{0.80\xspace}
\newcommand{\PubGapControlNothing}{27.4\%\xspace}
\newcommand{\PubGapControlCountryDisciplineRank}{47.0\%\xspace}
\newcommand{\PubGapControlCountryDisciplineRankLength}{12.4\%\xspace}

\newcommand{\PubGapControlDropout}{9.0\%\xspace}
\newcommand{\PubGapReducedByControl}{67\%\xspace}

\newcommand{\ImpactGap}{30\%\xspace}
\newcommand{\ImpactGapTopAuthor}{36\%\xspace}

\newcommand{\ImpactGapRecent}{34\%\xspace}

\newcommand{\ImpactGapControlNothing}{30.5\%\xspace}
\newcommand{\ImpactGapControlCountryDisciplineRank}{50.7\%\xspace}
\newcommand{\ImpactGapControlCountryDisciplineRankLength}{13.1\%\xspace}
\newcommand{\ImpactGapControlCountryDisciplineRankPub}{1.9\%\xspace}

\newcommand{\ImpactGapControlDropout}{12.1\%\xspace}

\newcommand{\ProdMale}{1.33\xspace}
\newcommand{\ProdFemale}{1.32\xspace}
\newcommand{\ProdGap}{0.9\%\xspace}
\newcommand{\ProdGapPValue}{10^{-9}\xspace}
\newcommand{\ProdGapTopAuthor}{4\%\xspace}

\newcommand{\LengthMale}{11.0\xspace}
\newcommand{\LengthFemale}{9.3\xspace}

\newcommand{\LengthGapChemistry}{19.2\%\xspace}

\newcommand{\LengthGapAppliedphysics}{2.5\%\xspace}

\newcommand{\CollabGapControlCountryDisciplineRank}{35.8\%\xspace}

\newcommand{\CollabGapControlCountryDisciplineRankPub}{4.1\%\xspace}

\newcommand{\CollabMaleControlCountryDisciplineRank}{36.6\xspace}

\newcommand{\CollabFemaleControlCountryDisciplineRank}{23.5\xspace}

\newcommand{\DropoutMale}{9.0\%\xspace}
\newcommand{\DropoutFemale}{10.8\%\xspace}
\newcommand{\DropoutGap}{19.5\%\xspace}
\newcommand{\SizeControlCountryDisciplineRank}{32,782\xspace}
\newcommand{\SizeControlCountryDisciplineRankLength}{25,033\xspace}

\newcommand{\RunsControlCountryDisciplineRankLength}{50\xspace}

\begin{document}

\begin{center}
\Huge{Historical comparison of gender inequality in scientific careers across countries and disciplines}\\
%\Huge{Quantifying gender inequality in scientific careers across countries and disciplines}\\
\vspace{1cm}
\large
{Junming Huang,$^{1,2,3,\ast}$ Alexander J.\ Gates,$^{1,\ast}$ Roberta Sinatra$^{4}$, \\ Albert-L{\'a}szl{\'o} Barab{\'a}si$^{1,5,6,\dagger}$\\
\vspace{0.5cm}
\small{$^{1}$Center for Complex Network Research, Northeastern University, Boston, Massachusetts 02115, USA}\\
\small{$^{2}$CompleX Lab, School of Computer Science and Engineering, University of Electronic Science and Technology of China, Chengdu 611731, China}\\
\small{$^{3}$Paul and Marcia Wythes Center on Contemporary China, Princeton University, Princeton, New Jersey 08540, USA}\\
\small{$^{4}$Department of Computer Science, IT University of Copenhagen, Copenhagen 2300, Denmark}\\
\small{$^{5}$Center for Cancer Systems Biology, Dana-Farber Cancer Institute, Boston, Massachusetts 02115, USA}\\
\small{$^{6}$Department of Medicine, Brigham and Women's Hospital, Harvard Medical School, Boston, Massachusetts 02115, USA}\\
\vspace{0.3cm}
\small{$^{\ast}$These authors contributed equally to this work.}\\
\small{$^{\dagger}$To whom correspondence should be addressed: \href{mailto:a.barabasi@northeastern.edu}{a.barabasi@northeastern.edu}}
}

\pagebreak
\setstretch{1.8}
%TC:break Abstract

\begin{sciabstract}
\normalsize
There is extensive, yet fragmented, evidence of gender differences in academia suggesting that women are under-represented in most scientific disciplines, publish fewer articles throughout a career, and their work acquires fewer citations.
Here, we offer a comprehensive picture of longitudinal gender discrepancies in performance through a bibliometric analysis of academic careers by reconstructing the complete publication history of over \PopulationApprox gender-identified authors whose publishing career ended between \YearStart and \YearEnd, covering \NumCountries countries and \NumDisciplines disciplines.
We find that, paradoxically, the increase of participation of women in science over the past 60 years was accompanied by an increase of gender differences in both productivity and impact.
Most surprisingly though, we uncover two gender invariants, finding that men and women publish at a comparable annual rate and have equivalent career-wise impact for the same size body of work.
Finally, we demonstrate that differences in dropout rates and career length explain a large portion of the reported career-wise differences in productivity and impact.
This comprehensive picture of gender inequality in academia can help rephrase the conversation around the sustainability of women's careers in academia, with important consequences for institutions and policy makers.

\end{sciabstract}

\end{center}

\pagebreak

%TC:break Body

\setstretch{2.5}

\label{introduction}
Gender differences in academia, captured by disparities in the number of female and male authors, their productivity, citations, recognition, and salary, are well documented across all disciplines and countries\cite{Ley2008gendergapgrant,Lariviere2013globalgender, Shen2013mindgendergap, Lincoln2012matilda, Holman2018closinggendergap, zippel2017review, Elseview2017gender,NAS2007beyond}.
The epitome of gender disparity is the ``productivity puzzle''\cite{Cole1984productivitypuzzle,Long1990origins,Long1992sexdifferenceproductivity,Xie1998sexdifferenceproductivity,Fox2017gender}---the persistent evidence that men publish more than women over the course of their career, which has inspired a plethora of possible explanations\cite{Abramo2009genderproductivity,Lariviere2011sexdifferences,Xie2003womenscience}, from differences in family responsibilities\cite{Carr1998children,Stack2004childrengap,Fox2005genderfamily}, to career absences\cite{Cameron2016genderbiasedmetrics}, resource allocation\cite{Duch2012genderresource}, the role of peer-review\cite{Borsuk2009namegender}, collaboration\cite{Jadidi2017genderdisparity,Uhly2015internationalcollaboration}, academic rank\cite{vandenBesselaar2017viciouscircles}, specialization\cite{Leahey2006genderproductivity}, and work climate\cite{Bronstein1998genderwork}. % add references
However, the deep inter-relatedness of these factors has limited our ability to differentiate the causes from the consequences of the productivity puzzle, complicating the scientific community's ability to enact effective policies to address it.

%Add here studies on organizational factors - "work climate," "accommodations to solve work-family conflicts" see comments

A key methodological obstacle has been the difficulty to reconstruct full scientific careers for scientists of both genders across the diverse academic population.
Consequently, much of the available evidence on gender disparity is based on case studies limited to subsets of active scientists in specific countries, disciplines, or institutions, making it difficult to compare and generalize the finding to all of science.
A further complication arises from the heavy-tailed nature of academia: a disproportionately small number of authors produce a large fraction of the publications and receive the majority of the citations\cite{Fortunato2018scisci}, an effect that is exacerbated in small sample sizes \cite{clauset2009power}.
To truly understand the roots of the gender gap, we need to survey the whole longitudinal, disciplinary, and geographical landscape, which is possible only if we capture complete careers for all scientists across disciplinary and national boundaries.

Here, we reconstructed the full career of \PopulationRaw scientists from their publication record in the Web of Science (WoS) database between 1900 and 2016.
By deploying a state-of-the-art method for gender identification (SI \ref{si:gender-assignment}), we identified the gender of over 3 million authors (\PopulationRawFemale female and \PopulationRawMale male) spanning \NumCountries countries and \NumDisciplines major disciplines (SI \ref{si:reorganize-discipline}).
We then focused on \Population scientists whose publishing careers ended between \YearStart and \YearEnd (Sections \ref{sec:si:data} and \ref{si:datasummary}), allowing us to systematically compare complete male and female careers.
To demonstrate the robustness of our findings to database bias and author disambiguation errors, we independently replicated our results in two additional datasets: the Microsoft Academic Graph\cite{mag} and the Digital Bibliography \& Library Project (DBLP), each utilizing different criteria for publication inclusion and methodologies for career reconstruction (SI \ref{sec:si:data} and \ref{sec:si:relicated}).
To our knowledge, our efforts constitute the most extensive attempt to date to quantify the gender gap in STEM publications and citations, offering a longitudinal, career-wise perspective across national and disciplinary boundaries.

\begin{figure*}[!p]
\begin{center}
	\includegraphics[height = 16cm]{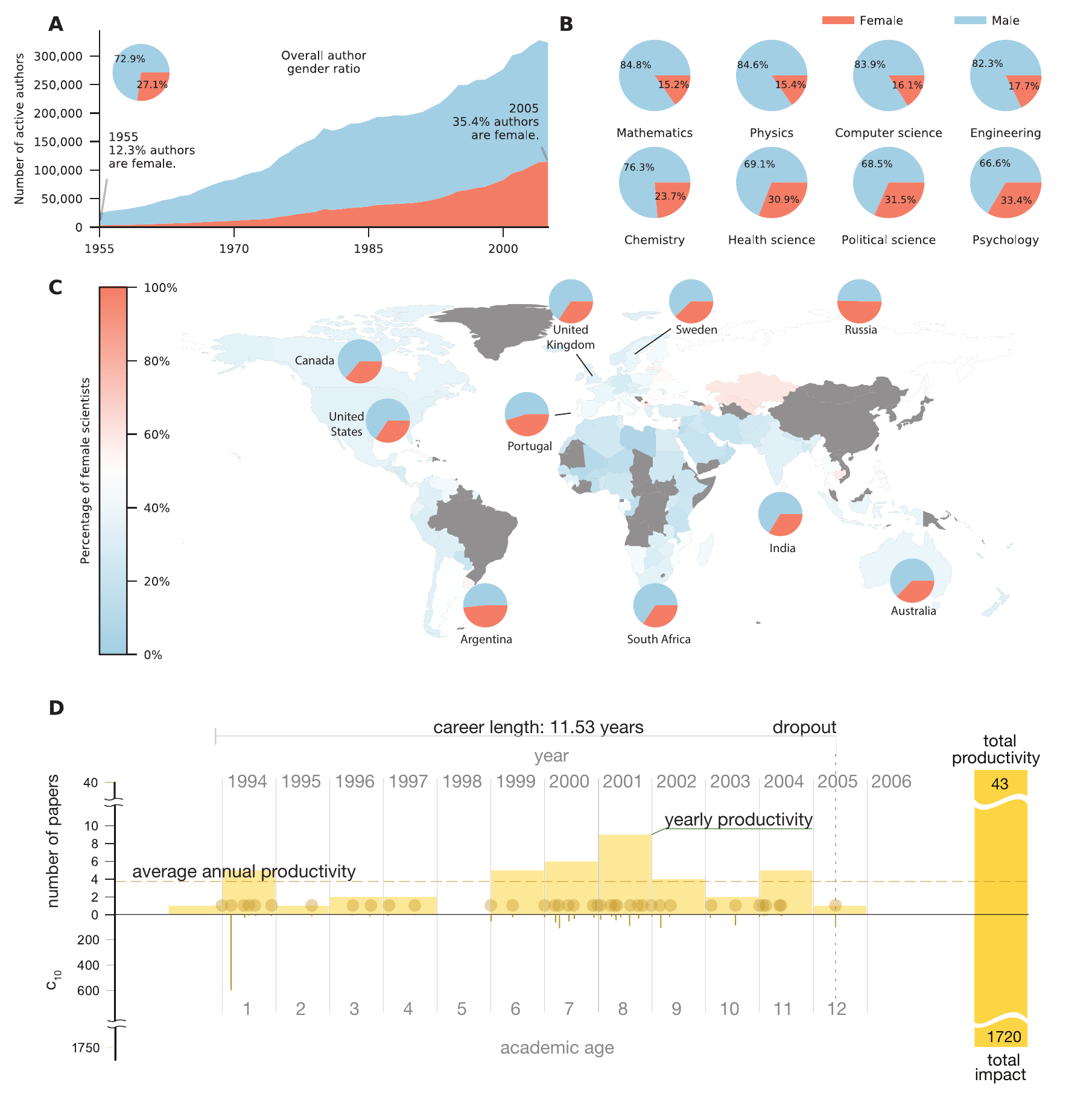}
	\caption{
	\textbf{Gender imbalance since \YearStart.}
	\textbf{A}, The number of active female (orange) and male (blue) authors over time and (inset) the total proportions of authors.
	\textbf{B},\textbf{C}, The proportion of female authors in several \textbf{B}, disciplines; and \textbf{C}, countries; for the full list see SI, Tables \ref{tab:discipline-performance} \& \ref{tab:country-performance}.
	\textbf{D}, The academic career of a scientist is characterized by his or her temporal publication record.
	For each publication we identify the date (gold dot) and number of citations after $10$ years $c_{10}$ (gold line, lower).
	The aggregation by year provides the yearly productivity (light gold bars), while the aggregation over the entire career yields the total productivity (solid yellow bar, right) and total impact (solid yellow bar, right).
	Career length is calculated as the time between the first and last publication, the annual productivity (dashed gold line) represents the average yearly productivity.
	An author drops out from our data when he published his last article.
}
 \label{fig:populationgap}
 \end{center}
\end{figure*}

Across all years and disciplines, women account for \FemalePercentage of authors, a number that hides important trends: while in \YearStart women represented only \FemalePercentageEarly of all active authors, that fraction steadily increased over the last century, reaching \FemalePercentageRecent by 2005 (\figref{fig:populationgap}{a}).
Yet, these aggregate numbers hide considerable disciplinary differences, as the fraction of women is as low as \FemalePercentageMathematics in math, physics and computer science, and reaches \FemalePercentagePsychology in psychology (\figref{fig:populationgap}{b}).
We also observe significant variations by country, finding that the proportion of female scientists can be as low as \FemalePercentageGermany in Germany, and reaches parity with \FemalePercentageRussia in Russia (\figref{fig:populationgap}{c}).

\begin{figure*}[!p]
\begin{center}
    \includegraphics[width = 16cm]{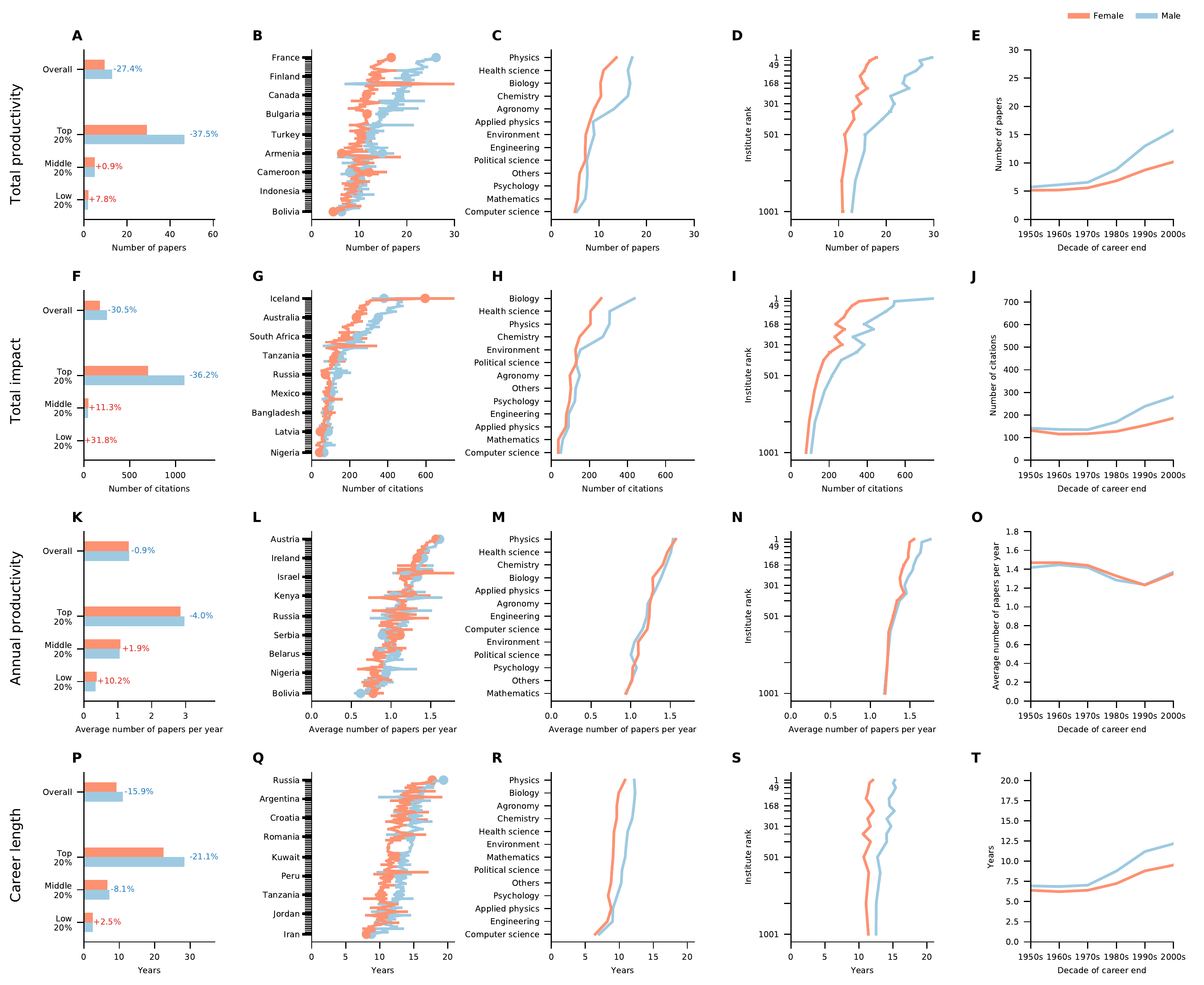}
	\caption{ \textbf{Gender gap in scientific publishing careers.} % add "publishing"
	The gender gap is quantified by the relative difference between the mean for male (blue) and female (orange) authors.
	In all cases the relative gender differences are statistically significant as established by the two-sided t-test, with p-values less than $10^{-4}$ unless otherwise stated (see SI \ref{sec:si:statsign} for test statistics).
	\textbf{A}-\textbf{E}, Total productivity broken down by: \textbf{A}, percentile; \textbf{B}, discipline; \textbf{C}, country; \textbf{D}, affiliation rank; and \textbf{E}, decade.
	The gender gap in productivity has been increasing from the 1950s to the 2000s.
	\textbf{F}-\textbf{J}, Total impact subdivided by: \textbf{F}, percentile; \textbf{G}, discipline; \textbf{H}, country; \textbf{I}, affiliation rank; and \textbf{J}, decade.
	\textbf{K}-\textbf{O}, Annual productivity is nearly identical for male and female authors when subdivided by: \textbf{K}, percentile; \textbf{L}, discipline; \textbf{M}, country; \textbf{N}, affiliation rank; and \textbf{O}, decade.
	\textbf{P}-\textbf{T}, Career length broken down by: \textbf{P}, percentile; \textbf{Q}, discipline; \textbf{R}, country; \textbf{S}, affiliation rank; and \textbf{T}, decade.
	}
 \label{fig:puzzle}
 \end{center}
\end{figure*}

The low proportion of women actively publishing in STEM captures only one aspect of gender inequality.
Equally important are the persistent productivity and impact differences between the genders (\figref{fig:populationgap}{d}).
We find that while on average male scientists publish \PubMale papers during their career, female authors publish only \PubFemale, resulting in a \PubGap gender gap in total productivity (\figref{fig:puzzle}{a}).
The difference is particularly pronounced among productive authors, as male authors in the top 20\% productivity bracket publish \PubGapTopAuthor more papers than female authors (\figref{fig:puzzle}{a}).
Interestingly, the gender disparity disappears for median productive authors (middle 20\%), and reverses for the authors in the bottom 20\%.
The gender gap in total productivity persists for all disciplines and all countries, with the exceptions of Cuba and Serbia (\figref{fig:puzzle}{b,c}).
We also observe a large gender gap in total productivity for the highest ranked affiliations (\figref{fig:puzzle}{d}, determined from the 2019 Times Higher Education World University Rankings, SI \ref{sec:si:affiliation}).

We measure the total impact during an academic career by the number of citations accrued 10 years after publication ($c_{10}$) by each paper published during a career (\figref{fig:populationgap}{d}), after removing self-citations and re-scaling to account for citation inflation\cite{Sinatra2016quantifyingimpact ,Wang2013longterm, Radicchi2008citationdist} (SI \ref{si:correct-citation}).
We find that male scientists receive \ImpactGap more citations for their publications than female scientists (\figref{fig:puzzle}{f}).
Once again, the total impact difference is the largest for high impact authors, and reverses for median and low impact authors: male authors in the top 20\% in career impact receive \ImpactGapTopAuthor more citations than their female counterparts.
The disparity in impact persists in all countries and disciplines, Iran and Serbia serving as the only exceptions (\figref{fig:populationgap}{g,h}), and can be found, to a lesser extent, across all affiliations regardless of affiliation rank (\figref{fig:populationgap}{i}).

Paradoxically, the gradual increase in the fraction of women in science\cite{Holman2018closinggendergap} (\figref{fig:populationgap}{a}), is accompanied by a steady increase in both the productivity and impact gender gaps (\figref{fig:puzzle}{e,j}).
The gender gap in total productivity rose from near \PubGapEarly in the 1950s, to a strong bias towards male productivity (\PubGapRecent gap) in the 2000s.
The gender gap in total impact actually switches from slightly more female impact in the 1950s to a \ImpactGapRecent gap favoring male authors in the same time frame.
These observations disrupt the conventional wisdom that academia can achieve gender equality simply by increasing the number of participating female authors.

In summary, despite recent attempts to level the playing field, men continue to outnumber women 2 to 1 in the scientific workforce, and, on average, have more productive careers and accumulate more impact.
These results confirm, using a unified methodology spanning most of science, previous observations in specific disciplines and countries\cite{Cole1984productivitypuzzle,Long1992sexdifferenceproductivity,Broder1993genderdifferenceeconomics,Xie1998sexdifferenceproductivity,Xie2003womenscience,Abramo2009starscientists,Lariviere2013globalgender,Maliniak2013gapinternationalrelations, West2013genderscholarlyauthorship}, and support in a quantitative manner the perception that global gender differences in academia is a universal phenomenon persisting in every STEM discipline and in most geographic regions.
Moreover, we find that the gender gaps in productivity and impact have increased significantly over the last 60 years.
The universality of the phenomenon prompts us to ask:
What characteristics of academic careers drive the observed gender-based differences in total productivity and impact?

As total productivity and impact over a career represent a convolution of annual productivity and career length, to identify the roots of the gender gap, we must separate these two factors.
Traditionally, the difficulty of reconstructing full careers has limited the study of annual productivity to a small subset of authors, or to career patterns observable during a fixed time frame\cite{Rorstad2015publicationrate,Symonds2006genderdifferences,Arensbergen2012persistingdifferences,Kaminski2012survivalfaculty,Box-Steffensmeier2015facultyretention,way2016gender,Way2017productivitytrajectory,Hechtman2018nihfunding}.
Access to the full career data allows us to decompose each author's total productivity into his/her annual productivity and career length, defined as the time span between a scientist's first and last publication (\figref{fig:populationgap}{d}, and SI, \ref{sec:si:career}).
We find that the annual productivity differences between men and women are negligible: female authors publish on average \ProdMale papers per year, while male authors publish on average \ProdFemale, a difference, that while statistically significant, is considerably smaller than other gender disparities (\ProdGap, p-value $<\ProdGapPValue$, \figref{fig:puzzle}{k}).
This result is observed in all countries, and disciplines (\figref{fig:puzzle}{l,m}) and we replicated it in all three datasets (SI, \ref{sec:si:relicated}).
The gender difference in annual productivity is small even among the most productive authors (\ProdGapTopAuthor for the top 20\%), and is reversed for authors of median and low productivity.

%This is such an important finding - are these gender differences significant for the median and low productivity authors?

The average annual productivity of scientists has slightly decreased over time, yet, there is consistently no fundamental difference between the genders (\figref{fig:puzzle}{o}).
In other words, when it comes to the number of publications per year, female and male authors are largely indistinguishable, representing the first gender invariant quantity in performance metrics.
As we show next, this invariant, our key result, helps us probe the possible roots of the observed gender gaps.

% This is such an important finding. But these last 2 sentences use weird language. Reformulate...

\begin{figure*}[!p]
\begin{center}
    \includegraphics[height = 17cm]{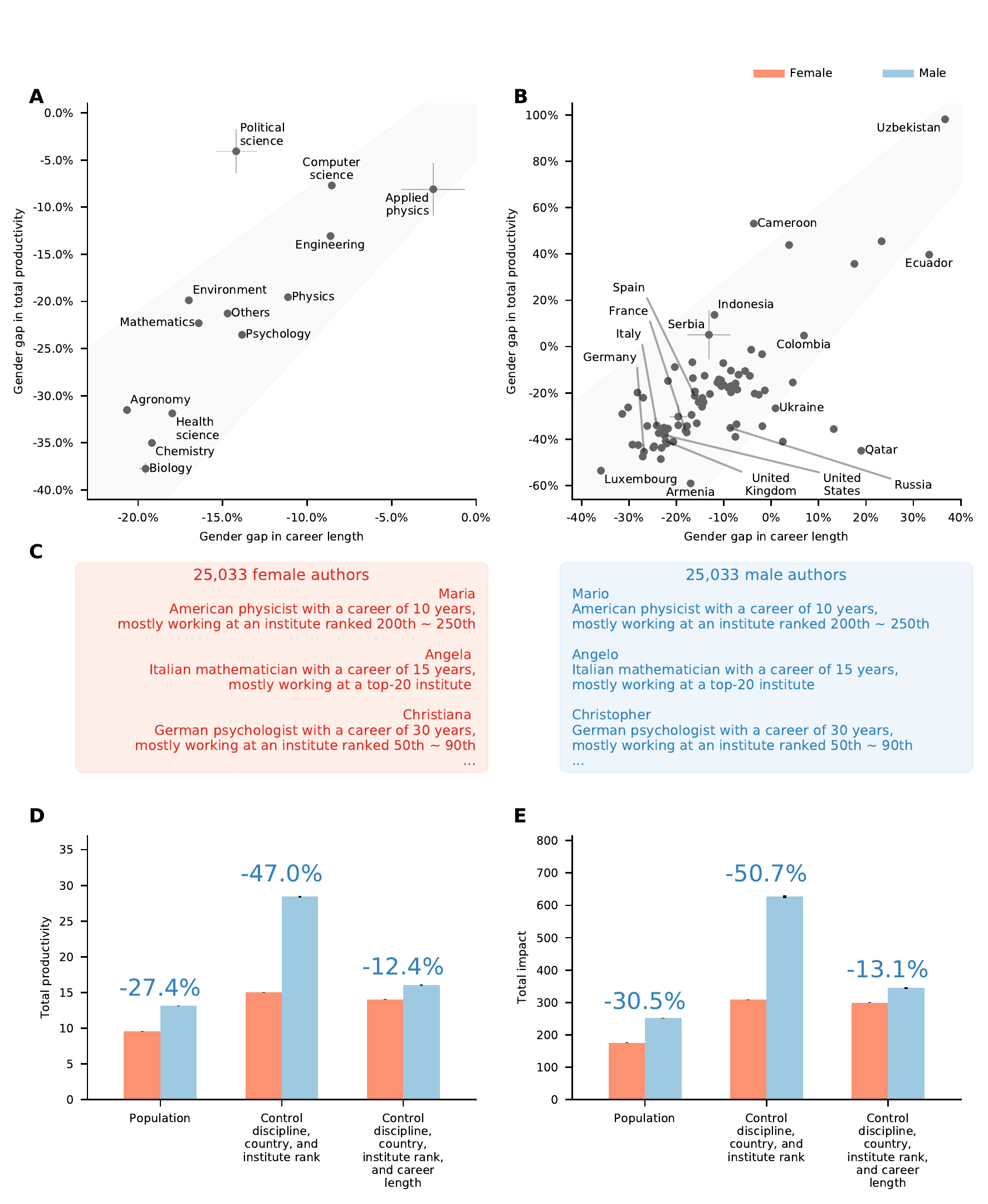}
	\caption{ \textbf{Controlling for career length.}
	\textbf{A},\textbf{B}, The gender gap in career length strongly correlates with the productivity gap across \textbf{A}, disciplines (Pearson correlation \PubGapDisciplinePearson) and \textbf{B}, countries (Pearson correlation \PubGapCountryPearson).
	\textbf{C}, In a matching experiment, equal samples are constructed by matching every female author with a male author having an identical discipline, country, and career length.
	\textbf{D}, The average productivity provided by the matching experiment for career length compared to the population; the gender gap is reduced from \PubGapControlNothing in the population to \PubGapControlCountryDisciplineRankLength in the matched samples.
	\textbf{E}, The average impact provided by the matching experiment for career length compared to the original unmatched sample.
	Where visible, error bars denote one std.
	}
 \label{fig:careerlength}
 \end{center}
\end{figure*}

The comparable annual productivity of male and female scientists suggests that the large gender gap in total career productivity is determined by differences in career length.
To test if this is the case, we measured the career length (time between first and last publication, \figref{fig:populationgap}{d}) of each scientist in the database, finding that, on average, male authors reach an academic age of \LengthMale years before ceasing to publish, while the average terminal academic age of female authors is only \LengthFemale years (\figref{fig:puzzle}{p}).
This gap persists when authors are grouped by either discipline, country, or affiliation (\figref{fig:puzzle}{q,r,s}), and has been increasing over the past 60 years (\figref{fig:puzzle}{t}).
Taken together, \figref{fig:puzzle}{k,t} suggest that a significant fraction of the variation in total productivity is rooted in variations in career lengths.
This conclusion is supported by a strong correlation between the career length gap and the career-wise productivity gap when we subdivide scientists by discipline (\figref{fig:careerlength}{a}, Pearson correlation \PubGapDisciplinePearson) and country (\figref{fig:careerlength}{b}, Pearson correlation \PubGapCountryPearson).
For example, the gender gap in career length is smallest in applied physics (\LengthGapAppliedphysics), as so is the gender gap in total productivity (\PubGapAppliedphysics).
In contrast, in biology and chemistry, men have \LengthGapChemistry longer careers on average, resulting in a total productivity gender gap that exceeds \PubGapChemistry.

Given the largely indistinguishable annual productivity patterns, we next ask how much of the total productivity and the total impact gender gaps observed above (\figref{fig:puzzle}{a,f}) could be explained by the variation in career length.
For this, we perform a matching experiment designed to eliminate the gender gaps in career length.
In the first population, for each female scientist, we select a male scientist from the same country and discipline, and whose primary affiliation is ranked approximately the same (\figref{fig:careerlength}{c}, and SI \ref{sec:si:matching}).
In this matched population, the gender gap in total productivity increases significantly, from \PubGapControlNothing to \PubGapControlCountryDisciplineRank, and the gender gap in total impact increases from \ImpactGapControlNothing to \ImpactGapControlCountryDisciplineRank.
This increase in both the total career statistics and gender gaps occurs because access to country and affiliation information is biased towards more recent and senior scientists.
We then constructed a second matched population, as a subset of the first, in which each female scientist is matched to a male scientist from the same country, discipline, affiliation rank, and with exactly the same career length.
In these career length matched samples the gender gap in total productivity reduces from \PubGapControlCountryDisciplineRank to \PubGapControlCountryDisciplineRankLength (\figref{fig:careerlength}{d}).
Furthermore, the gender gap in the total impact is also reduced from \ImpactGapControlCountryDisciplineRank to \ImpactGapControlCountryDisciplineRankLength (\figref{fig:careerlength}{e}).
By matching pairs of authors based on observable confounding variables, such as their country, discipline, and affiliation rank, we mitigate the influence of these variables on the gender gaps.
While matching cannot rule out that gender differences are influenced by unmatched variables that are unobserved here, the significant decrease in the productivity and impact gender gaps when we control for career length suggests that publication career length is a significant correlate of gender differences in academia.

Thus far, our analysis has correlated the career-wise gender gaps to systematic differences in career lengths, prompting us to ask: Does the persistent gender differences in the cessation of academic publishing drive the career-wise gender gaps?
While a full assessment of causality would require us to conduct a controlled intervention on the academic population (a wholly unfeasible scenario), our matching experiments suggest a counterfactual experiment to identify, at the population level, the average causal effect of shorter careers on total productivity and impact.

\begin{figure*}[!p]
\begin{center}
    \includegraphics[width = 16cm]{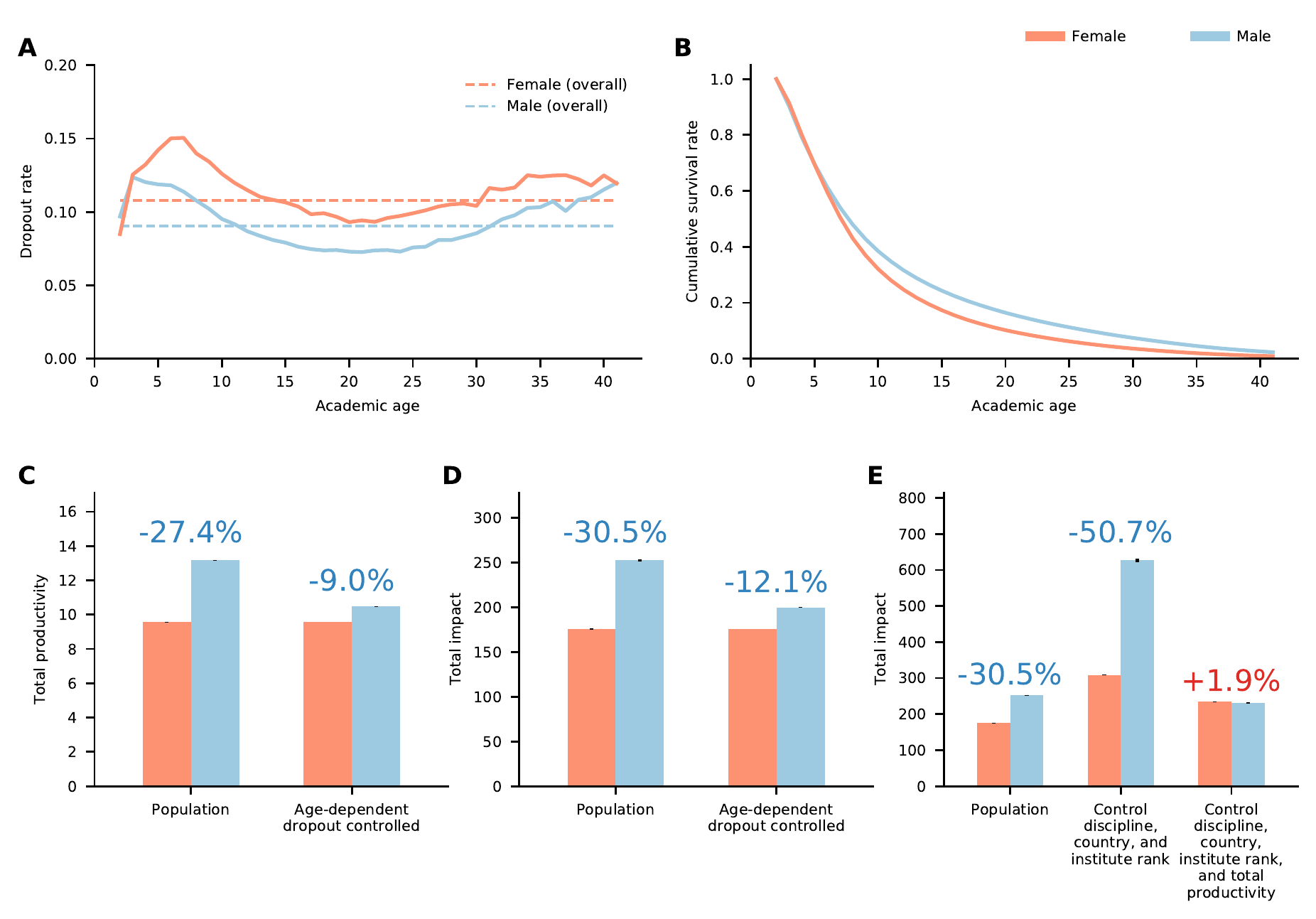}
	\caption{ \textbf{Author's age-dependent dropout rate.}
	\textbf{A}, Dropout rate for male (blue) and female (orange) authors over their academic ages.
	\textbf{B}, The cumulative survival rate for male and female authors over their academic ages.
	\textbf{C},\textbf{D}, The effect of controlling for the age-dependent dropout rate on the gender gaps in \textbf{C}, total productivity and \textbf{D}, impact.
	\textbf{E}, The total impact gap is eliminated in the matched sample based on total productivity.
	%
	%Where visible, shaded regions and error bars denote one std.
    }
 \label{fig:dropout}
 \end{center}
\end{figure*}

To address the factors governing the end of a publishing career, we calculated the dropout rate, defined as the yearly fraction of authors in the population who have just published their last paper\cite{Kaminski2012survivalfaculty, Milojevic2018demographics}.
We find that on average \DropoutMale of active male scientists stop publishing each year, while the yearly dropout rate for women is nearly \DropoutFemale (\figref{fig:dropout}{a}).
In other words, each year women scientists have a \DropoutGap higher risk to leave academia than male scientists, giving male authors a major cumulative advantage over time.
Moreover, this observation demonstrates that the dropout gap is not limited to junior researchers, but persists at similar rates throughout scientific careers.

The average causal effect of this differential attrition is demonstrated through a counter-factual experiment in which we shorten the careers of male authors to simulate dropout rates matching their female counterparts at the same career stage (\figref{fig:dropout}{c,d}, and SI \ref{sec:si:dropoutcontrol}).
We find that under similar dropout rates, the differences in total productivity and total impact reduce by roughly two thirds, namely from \PubGapControlNothing to \PubGapControlDropout and from \ImpactGapControlNothing to \ImpactGapControlDropout respectively.
This result, combined with our previous matching experiment (\figref{fig:careerlength}{d,e}), suggests that the difference in dropout rates is a key factor in the observed total productivity and impact differences, accounting for about \PubGapReducedByControl of the productivity and impact gaps.
Yet, the differential dropout rates do not account for the whole effect, suggesting that auxiliary disruptive effects, from perception of talent to resource allocation\cite{Lariviere2011sexdifferences, Duch2012genderresource}, may also play a potential role.

The reduction of the gender gaps in both total productivity and total impact by similar amounts suggests that total impact, being the summation over individual articles, may be primarily dependent on productivity \cite{Lariviere2011sexdifferences}.
To test this hypothesis, we conducted a final matching experiment in which we selected a male author from the same country, discipline, approximately the same affiliation rank, and with exactly the same number of total publications as each female author (SI \ref{sec:si:prodmatching}).
In these matched samples the gender gap in the total impact is completely eliminated, dropping from \ImpactGapControlCountryDisciplineRank in favor of male authors, to \ImpactGapControlCountryDisciplineRankPub in favor of female authors (\figref{fig:dropout}{e}).
This reveals a second gender invariant quantity---there is no discernible difference in impact between male and female scientists for the same size body of work.
This second gender invariant reinforces our main finding that it is career length differences which drive the total productivity gap, that consequently drive the impact gender gap in academia.
Interestingly, controlling for productivity similarly flips the gender gap in the total number of collaborators throughout a career (SI \ref{si:sec:collab}).

%\section*{Summary and Discussion}
%
%
%
Our ability to reconstruct the full careers of scientists allowed us to confirm the differences in total productivity and impact between female and male scientists across disciplines and countries since \YearStart.
We showed that the gradual increase in the fraction of women in STEM was accompanied by an increase in the gender disparities in productivity and impact.
It is particularly troubling that the gender gap is the most pronounced among the highly productive authors---those that train the new generations of scientists and serve as role models for them.
Yet, we also found two gender invariants, revealing that active female and male scientists have largely indistinguishable yearly performance, and receive a comparable number of citations for the same size body of work.
These gender-invariant quantities allowed us to show that a large portion of the observed gender gaps are rooted in gender-specific dropout rates and the subsequent gender-gaps in career length and total productivity.
This finding suggests that we must rephrase the conversation about gender inequality around the sustainability of woman's careers in academia, with important administrative and policy implications \cite{Xie2003womenscience,Etzkowitz2000athena,Ceci2011understandingunderrepresentation,Maliniak2013gapinternationalrelations,Sheltzer2014elitehire,Williams2015nationhiring,Nielsen2017genderdiversity,Cech2019changingcareer}.

It is often argued that in order to reduce the gender gap, the scientific community must make efforts to nurture junior female researchers.
We find, however, that the academic system is loosing women at a higher rate at every stage of their careers, suggesting that focusing on junior scientists alone may not be sufficient to reduce the observed career-wise gender imbalance.
The cumulative impact of this career-wide effect dramatically increases the gender disparity for senior mentors in academia, perpetuating the cycle of lower retention and advancement of female faculty\cite{Long1990origins,NationalResearchCouncil2010,Martinez2017feelingivorytower,Cech2019changingcareer}.

Our focus on closed careers limited our study to careers that ended by \YearEnd, eliminating currently active careers.
Therefore, further work is needed to detect the impact of recent efforts by many institutions and funding agencies to support the participation of women and minorities\cite{Arensbergen2012persistingdifferences, Stewart2018inclusive}.
Our analysis of all careers and the factors that dominate the gender gap could offer a base line for such experimental studies in the future.
At the same time, our work suggests the importance of temporal controls for studying academic careers, and in particular, gender inequality in academia.

It is important to emphasize that the end of a publishing career does not always imply an end of an academic career; authors who stopped publishing often retain teaching or administrative duties, or conduct productive research in industry or governmental positions, with less pressure to communicate their findings through research publications.
Scientific publications represent only one of the possible academic outputs; in some academic disciplines books and patents are equally important, and all three of our data sources (WoS, MAG and DBLP) tend to over-represent STEM and English language publications\cite{mongeon2016journalcoverage}, thereby possibly biasing our analysis.
Furthermore, our bibliometric approach can draw deep insight into the large-scale statistical patterns reflecting gender differences, yet we cannot observe and test potential variation in the organizational context and resources available to individual researchers\cite{Wennerc1997nepotism, Fox2017gender}.
However, our results do suggest important consequences for the organizational structures within academic departments.
Namely, we find that a key component of the gender gaps in productivity and impact may not be rooted in gender-specific processes through which academics conduct research and contribute publications, but by the gender-specific sustainability of that effort over the course of an entire academic career.

%TC:break EndMatter

\singlespace

\section*{Acknowledgements}
Special thanks to Alice Grishchenko for help with the visualizations.
Also, thanks to the wonderful research community at the CCNR, and in particular Yasamin Khorramzadeh, for helpful discussions, and to Kathrin Zippel at Northeastern University for valuable suggestions.
 A.J.G. and A.-L.B.\ were supported in part by the Templeton Foundation, contract \#61066.
 J.H. and A.-L.B. were supported in part by the Defense Advanced Research Projects Agency (DARPA), contract DARPA-BAA-15-39.
 R.S. acknowledges support from Air Force Office of Scientific Research grants FA9550-15-1-0077 and FA9550-15-1-0364.
 The authors declare that they have no competing financial interests.

\section*{Data \& Code Availability}
The DBLP and MAG are publicly available from their source websites (see SI).
Other related and relevant data and code are available from the corresponding author upon request.

\section*{Author contributions}
J.H., A.J.G., R.S., and AL.B. collaboratively conceived and designed the study, and drafted, revised, and edited the manuscript.  J.H. and A.J.G. analyzed the data and ran all simulations.

\section*{Supplementary Materials}
Materials and Methods\\
Table S1 - S8\\
Fig S1 - S8\\

\pagebreak

% 17.8cm, 11.4cm
%%%%%%%%%%%%%%%%%%%%%%
%%%%%%%%%%%%%%%%%%%%%%%%

% 17.8cm, 11.4cm
%%%%%%%%%%%%%%%%%%%%%%
%%%%%%%%%%%%%%%%%%%%%%%%

% 17.8cm, 11.4cm
%%%%%%%%%%%%%%%%%%%%%%
%%%%%%%%%%%%%%%%%%%%%%%%

% 17.8cm, 11.4cm
%%%%%%%%%%%%%%%%%%%%%%
%%%%%%%%%%%%%%%%%%%%%%%%

\clearpage

%TC:break SI

{\huge \textbf{Supplemental Information}}\\

{\LARGE Historical comparison of gender inequality in scientific careers across countries and disciplines}\\

% Author and supervisor
Junming Huang$^{1,2,3,*}$, Alexander J.\ Gates$^{1,*}$, Roberta Sinatra$^{4}$, and Albert-L{\'a}szl{\'o} Barab{\'a}si$^{1,5,6,\dagger}$

$^{1}$Center for Complex Network Research, Northeastern University, Boston, Massachusetts 02115, USA

$^{2}$CompleX Lab, School of Computer Science and Engineering, University of Electronic Science and Technology of China, Chengdu 611731, China

$^{3}$Paul and Marcia Wythes Center on Contemporary China, Princeton University, Princeton, New Jersey 08540, USA

$^{4}$Department of Computer Science, IT University of Copenhagen, Copenhagen 2300, Denmark

$^{5}$Center for Cancer Systems Biology, Dana-Farber Cancer Institute, Boston, Massachusetts 02115, USA

$^{6}$Department of Medicine, Brigham and Women's Hospital, Harvard Medical School, Boston, Massachusetts 02115, USA

* These authors contributed equally to this work.

$\dagger$ To whom correspondence should be addressed: \href{mailto:a.barabasi@northeastern.edu}{a.barabasi@northeastern.edu}

\tableofcontents

\pagebreak

\doublespace
\beginsupplement

%\newpage

\section{Data sets}\label{sec:si:data}

\subsection{Web of Science}
    The primary source of publication data for this project is the Clarivate Analytics' Web of Science Core Collection (WoS) database, covering the Science Citation Index Expanded and the Social Sciences Citation Index.
    We considered all articles, reviews, and letters published between 1900 to 2016, and we excluded all other types of documents (e.g.\ editorials, letters to the editor, and book reviews), published that are generally not peer-reviewed.
    In total, we consider the publication history of \PopulationRaw authors who contributed a total of 101,961,318 authorships to 53,788,499 publications.
    Additionally, we extracted the citation history for all publications, resulting in 694,439,758 citation relationships.

    The WoS dataset assigns each article to at least one scientific discipline in a three-layer hierarchy of 153 disciplines.
    For example, a paper is assigned to ``Science \& Technology'' (top layer), ``Life Sciences \& Biomedicine'' (middle layer) and ``Biophysics'' (leaf layer).
    The assignment is primarily based on each publication's journal information, but a select few multidisciplinary journals (e.g.\ \textit{Nature} and \textit{Science}) provide article-specific categories.
    For our purposes, the 153 disciplines in the leaf layer are too fine grained, while the other two layers do not provide a detailed enough classification.
    Therefore, we grouped the leaf layer categories into a coarser partition as described in Section \ref{si:reorganize-discipline}.

\subsection{Microsoft Academic Graph}
    The Microsoft Academic Graph (MAG) is a comprehensive index of scientific publications in both journals and conferences\cite{mag}.
    In November 2017, we downloaded 77,642,549 publications through the authorized API, freely provided by Microsoft Research available at \url{https://www.microsoft.com/en-us/research/project/microsoft-academic-graph/}.
    These publications were produced by 88,223,538 authors who contributed a total of 211,897,481 authorships.

\subsection{DBLP}
    The DBLP Computer Science Bibliography contains 4,181,940 publications from computer science journals and conference proceedings (downloaded June 5th, 2018, \url{https://dblp.uni-trier.de}).
    We consider all articles, review articles, proceedings, book chapters, and dissertations published between 1970 and 2010, and exclude all other types of documents (e.g.\ webpages and notes), that are generally not peer-reviewed.
    These publications were produced by 2,129,492 authors who contributed a total of 12,090,783 authorships.

\section{Data pre-processing}

\subsection{Identifying scientific careers}\label{sec:si:career}

    While the problem of name disambiguation for scientific publications is notoriously difficult, the scientific community has recognized several disambiguation procedures that effectively capture scientific careers.
    Here, to demonstrate the robustness of our results to database bias and author disambiguation errors, we replicated our analysis in several databases, each with its own strengths and weakness.
    %
    %The most simplistic approach matches authors purely based on first, last and middle names \cite{Milojevic2018demographics,Sekara2018chaperone}.
    %
    All three of the data sets we used (WoS, MAG, and DBLP) maintain unique author identifiers based on a different name disambiguation procedure.
    The WoS and MAG use their own proprietary algorithms which have been successfully used to study scientific careers (for example, see WoS\cite{Liu2018hotstreaks}, and MAG\cite{AlShebli2018ethnicdiversity}).
    While the specifics of the algorithms are not available, it is reasonable to assume that both algorithms are on par, if not far better than prevailing methods developed by independent academic groups.
    For instance, the MAG processes online CVs and Wikipedia profiles to associate individual authors with their papers.
    Additionally, both algorithms incorporate the self-curated career profiles provided by the Open Researcher and Contributor ID (ORCID).
    On the other hand, the DBLP name disambiguation is based on a unique identifier assigned to authors when manuscripts are submitted to registered Computer Science conferences or journals.
    Thus, the DBLP database has arguably the most reliable name disambiguation available in a bibliometric database\cite{Reuther2006dblpnames}, and has also been used in several peer-reviewed studies to study scientific careers\cite{Jadidi2017genderdisparity, Way2017productivitytrajectory}.

    While many of the name disambiguation algorithms are able to reconstruct the careers for authors with European names, they often have difficulty disambiguating the careers of authors with Asian names.
    This, combined with the known issues inferring the gender of Asian names (see below), motivates us to adapt a conservative approach and exclude all researchers from China (mainland, Hong Kong, Macau, \& Taiwan), the Democratic People's Republic of Korea, Japan, Malaysia, the Republic of Korea, and Singapore.

    Critically, by replicating our study in three different databases, each with an independent method for name disambiguation, we argue that any possible errors resulting from misappropriated or missing publications are negligible.

\subsection{Career selection criteria}

    In order to study comprehensive scientific careers, we limit our analysis to authors that:
    (i) have authored at least two papers,
    (ii) their publication careers span more than one year (365 days),
    (iii) have an average annual publication rate of less than 20 papers per year,
    (iv) have published their last article on or before Dec 31st, 2010.
    Our main conclusions do not change if more stringent selection criteria or modified filters are used to select the subset of scientists.

\subsection{Country label}\label{sec:si:country}

    To facilitate the assignment of author gender (Section \ref{si:gender-assignment}) and analyze national variations in the gender gap, we associate each author to a single country as follows.
    In the WoS, many authorships are indexed along with an affiliation address, including an institution name, street address, city, zipcode and country.
    For each author, we identify all authorships with a known affiliation address and keep only the country of the affiliation.
    We then assign a country label to an author based on the most frequently occurring country of affiliation.
    This frequency-based method results in a country label for a total of 1,876,950 authors.

    We also considered an alternative method for country assignment in which the earliest country affiliation was used for each author.
    This second method disagrees with the frequency-based approach for only 58,576 ($3.12\%$) of authors, and does not qualitatively affect results.

    For the country-specific analysis, we disregard countries with less than 100 male or 100 female authors because the sample size is not sufficiently large to produce reliable statistics.
    This results in the following \NumCountries countries reported in country-specific analysis in the main manuscript: \CountryList.

\subsection{Affiliation rank}\label{sec:si:affiliation}

    It has been suggested that the author's primary affiliation contributes significantly towards the overall productivity\cite{Way2019prominence}.
    We collected the ranking information from The Times Higher Education World University Rankings 2019\footnote{\href{https://www.timeshighereducation.com/world-university-rankings/2019/world-ranking\#!/page/0/length/25/sort\_by/rank/sort\_order/asc/cols/stats}{https://www.timeshighereducation.com/world-university-rankings/2019/world-ranking\#!/page/0/length/25/sort\_by/rank/sort\_order/asc/cols/stats}, accessed May 2019}, a global ranking that indexes more than 1,250 universities.
    Then we associate authors with those universities by examining the affiliations in their publications.
    Considering university names could be spelled in multiple ways, such as abbreviations, we queried every affiliation name in the Web of Science publication data, as well as all university names in the Times Higher Education World University Rankings, with Google Maps to disambiguate those variations into unique university names.
    Each author is then assigned the rank of the highest ranked institute to which she or he is affiliated over the course of the career.
    Among 1,876,950 authors with at least one affiliation recorded, 1,296,995 authors have been aligned to an institute rank.

    %Rank data are collected from three mainstream sources: Times 2019, ARWU 2018, QS 2019. In each ranking, over 1,000 universities are ranked. I use Google Maps to do the name disambiguation on university names, and almost all ranked universities have been successfully aligned to an institute in our data. Usually an author is affiliated with multiple institutes, I calculate the highest rank or the weighted average rank as his/her rank. Therefore I've prepared a total of 6 variations of institute rank of each author. Below I report the results based on Times + highest rank

\subsection{Gender assignment}\label{si:gender-assignment}

    % method: first name, country, genderize, confidence
    %\subsubsection{Gender assignment}
    In the absence of gender information for authors in the WoS, MAG, and DBLP we infer author gender based on author name and country.
    Specifically, we used a commercially available service \emph{Genderize.io}\footnote{\href{https://genderize.io/}{https://genderize.io/}} which integrates publicly available census statistics to build a name database mapping a country-specific first name to a binary gender label.
    Due to a low accuracy of the gender assignment algorithm for Asian names, we excluded all researchers from China (mainland, Hong Kong, Macau, \& Taiwan), the Democratic People's Republic of Korea, Japan, Malaysia, the Republic of Korea, and Singapore.
    We also excluded researchers from Brazil. %Why?
    This gender assignment strategy has been successfully employed in several academic research projects\cite{Jadidi2017genderdisparity,Holman2018closinggendergap,AlShebli2018ethnicdiversity}

    % first name completion
\subsubsection{WoS and MAG authorship alignment}
    A practical challenge lies in the fact that the WoS dataset records the full first name of authors on most papers published after 2006, while most papers before 2006 authorships are recorded with initials only.
    Among a total of 7,817,639 authors in the Web of Science dataset, only 2,171,290 of them have the full first name recorded for at least one authorship.
    Therefore, we leveraged our access to multiple datasets to help complete the missing metadata from the papers.
    Specifically, we aligned papers in the WoS to MAG based on the following criteria:
    (a) both papers are published in the same year,
    (b) both papers have identical sets of author last names,
    (c) the two papers differ in title by no more than 25\%, estimated by the Levenshtein distance between two titles divided by the length of the WoS paper title.
    Such matches were found for 23,615,112 papers.
    We aligned authorships in each paper pair by comparing first initial and last name.
    For example, if a WoS paper records an author ``J. Smith'' and its matched paper in MAG records ``John Smith'', we complete the authorship ``J Smith'' with ``John Smith''.
    We skipped papers with multiple authors sharing the same last name.
    This procedure allowed us to complete the first name for additional 1,334,886 authors.

    Note that this procedure only filled in missing metadata at the level of individual papers.
    The alignment between WoS and MAG was not sued to infer an author's career.

\subsubsection{Gender label inference}
    Out of the 3,427,232 WoS authors with full first name, we successfully inferred the gender of 3,003,815 authors, including 2,146,926 male authors and 856,889 female authors.

    % evaluation
\subsubsection{Gender label accuracy}
    As reported in Karimi et al.\ (2016)~\cite{Karimi2016gendernames}, genderize.io achieves a minimum accuracy of 80\%.
    To assess the accuracy of the gender assignment process for our data, we compared the inferred gender labels of authors in the WoS with a ground truth benchmark dataset consisting of 2,000 male and female full names manually collected in Lariviere et al.\ (2013)~\cite{Lariviere2013globalgender}.
    Among the 1,512 author names that overlap with our dataset, 1,425 have inferred gender labels that agree with the ground truth, resulting in an accuracy of $94.25\%$.

\subsection{Citation count and normalization}\label{si:correct-citation}

    \subsubsection{Citations within Web of Science}
    We only count citations in which both the Citing paper and Cited paper appear within the WoS database.

    \subsubsection{Removing self-citations}
    It has previously been shown that male scientists are more likely to cite their own papers than female scientists\cite{King2017genderselfcite}.
    Therefore, in all measures of impact, we removed all self-citations based on the overlap between authorships in the citing paper and cited papers.
    We also replicated our analysis while keeping all self-citations and found no qualitative difference in our primary conclusions.

    \subsubsection{Citation normalization}
    Citation-based measures of impact are affected by two major problems:
    (1) citations follow different dynamics for different papers\cite{Wang2013longterm} and
    (2) the average number of citations changes over time\cite{Sinatra2016quantifyingimpact}.
    To overcome the first problem, we focused on the total number of citations each paper received within 10 years after its publication, $c_{10}$, as a measure of its scientific impact.
    We corrected for the second problem by normalizing the $c_{10}$ for each paper by the average $c_{10}$ of papers published in the same year, and multiplying by 12 (an arbitrary constant that does not quantitatively affect any of our analysis but restores the normalized citation count back to a realistic value).
    The resulting normalized $c_{10}$ score thus provides a consistent measure of impact across decades.

\subsection{Discipline hierarchy}\label{si:reorganize-discipline}
    % re-organize discipline tree, select 12 disciplines with most authors

    % define author discipline

    We used a classification of scientific fields as defined in Wikipedia\footnote{Last accessed August 2018.
    \href{https://en.wikipedia.org/wiki/Branches_of_science}{Branches of science (Wikipedia)},
    \href{https://en.wikipedia.org/wiki/Outline_of_natural_science}{Outline of natural science (Wikipedia)},
    \href{https://en.wikipedia.org/wiki/Outline_of_social_science}{Outline of social science (Wikipedia)},
    \href{https://en.wikipedia.org/wiki/Outline_of_applied_science}{Outline of applied science (Wikipedia)}
    }
    to re-organize 153 WoS categories into 75 disciplines.
    See~\ref{tab:reorganize-discipline} for the details of the mapping.

    Each paper is assigned one or more disciplines among the 75 Wikipedia disciplines based on its original WoS category label(s).
    3,117,710 (39.66\%) authors have all papers assigned to a single discipline, while the remaining 4,742,941 (60.34\%) authors are associated with at least two disciplines.
    For each author with multiple disciplines, we assign with a single discipline label as the most frequently occurring one.
    3,728,442 (78.61\%) of 4,742,941 authors with multiple disciplines have the most frequent discipline occurring in more than half of his/her papers.

    While some disciplines were associated with many authors (e.g.\ Heath Sciences has 584,628 authors), many were only associated with a few authors.
    Therefore, we limit the majority of our analysis to the top 12 disciplines based on total population: \textbf{Health Science, Biology, Chemistry, Engineering,Physics, Computer Science, Psychology, Agronomy, Mathematics, Environmental science, Political Science, Applied physics}.
    These 12 disciplines cover 90.3\% of the population.
    The remaining 9.7\% of the population are grouped into the 13th category \textbf{Others} containing 4 fields in Formal Sciences (Decision theory, Logic, Statistics, Systems theory),
    9 fields in Natural Sciences (Botany, Earth science, Ecology, Geology, Human biology, Meteorology, Oceanography, Space Science and Astronomy, Zoology),
    14 fields in Applied Sciences (Applied chemistry, Applied linguistics, Applied mathematics, Architecture, Computing technology, Education, Electronics, Energy storage, Energy technology, Forensic science, Management, Microtechnology, Military science, Spatial science),
    30 fields in Social Sciences (Anthropology, Business studies, Civics, Cognitive Science, Criminology, Cultural studies, Demography, Development studies, Economics, Education, Environmental studies, Gender and sexuality studies, Geography, Gerontology, Industrial relations, Information science, International studies, Law, Legal management, Library science, Linguistics, Management, Media studies, Paralegal studies, Planning, Public administration, Social work, Sociology, Sustainability studies, Sustainable development),
    5 fields in Arts and Humanities (Arts, History, Languages and literature, Philosophy, Theology), and one last field ``Unknown'' that we failed to map to any Wikipedia discipline.

\subsection{Data summary}
    \label{si:datasummary}
    After all data processing steps were completed, we consider \Population WoS authors (\PopulationMale male, \PopulationFemale female), contributing 18,750,502 authorships to 13,081,184 papers, across \NumDisciplines disciplines and \NumCountries countries.

\section{Indicators}

\subsection{Characterizing the scientific career}\label{si:metrics}

    \begin{enumerate}
        \item \textbf{Total productivity} of a scientist is defined as the total number of authorships published by a specific author.

        \item \textbf{Career Length} of a scientist is defined as the difference between the date of publication for their first and last publications.
        The career length is naturally found at the resolution of days, while in coarser scenarios we report career length in years by dividing by $365$ and rounding to the nearest integer.

        \item \textbf{Annual Productivity} of a scientist is calculated as the ratio of total productivity to career length, i.e., (the total number of papers) / (the days between the first and last publications / 365).

        \item \textbf{Total impact} is defined as the sum of normalized $c_{10}$ scores for each paper published by a specific author.

        \item \textbf{Academic Age} of a scientist counts the number of years since his/her first publication.
        For example, a scientist whose first publication was in 1991, will have an academic age of 5 in 1995. %Academic age is a one-based integer.

        \item \textbf{Dropout} of a scientist occurs when the scientist publishes their final paper recorded in the data.

    \end{enumerate}

\subsection{Characterizing the scientific population}
    \begin{enumerate}

        \item \textbf{Gender gap} is calculated for each indicator as the relative difference, i.e., the difference between the mean female and male values divided by the value of the male indicator.

        \item \textbf{Dropout rate} of a group of scientists (e.g., those at the same age etc.) is the proportion of scientists who dropout from the group in the next year.
        %
        %\agnote{Hazard function and Kaplan–Meier estimator?}

    \end{enumerate}

\section{Methods}

  \subsection{Statistical significance}\label{sec:si:statsign}

        For each measurement of scientific performance, we report the gender gap as the difference between the mean value for female and male scientists.
        Additionally, we compute the statistical significance of the gap using the unpaired two-tailed Welch's t-test to detect whether two samples with unequal size and unequal variance deviate from the null hypothesis that the two distributions (female and male) have the same mean.
        The corresponding p-values, indicating the statistical significance of the test, are reported in Tables~\ref{tab:discipline-performance},
        ~\ref{tab:country-performance},
        ~\ref{tab:start-performance},
        % ~\ref{tab:start-decade-performance},
        ~\ref{tab:end-performance},
        % ~\ref{tab:end-decade-performance}.

    % matched samples
    \subsection{Career length matching}\label{sec:si:matching}

        In order to assess the relationship between career length and total productivity, we conducted a matching experiment as follows.
        We first constructed a matched baseline population, in which, for each female author, we identified, without-replacement, a male author from the same country, discipline, and with approximately the same affiliation rank.
        If multiple male authors were found, we randomly selected one to match.
        This process consistently produced \SizeControlCountryDisciplineRank matched pairs.
        To account for the inherent randomness in this procedure, the experiment was replicated \RunsControlCountryDisciplineRankLength times, and the reported performance was averaged over all random trials.
       The standard deviation over the trials is near zero for both the productivity and impact gaps.
        For matches based on affiliation, we binned the institutions by rank into 15 equal volume bins, and matched within the same bin; no significant difference occurs for other choices of the affiliation binning.
        We then created our second experimental population, as a subset of the first, in which we matched each female author to a male author from the same country, discipline, with approximately the same affiliation rank, and with exactly the same career length.
        This process consistently produced \SizeControlCountryDisciplineRankLength matched pairs.

\begin{figure*}[!p]
        \begin{center}
            \includegraphics{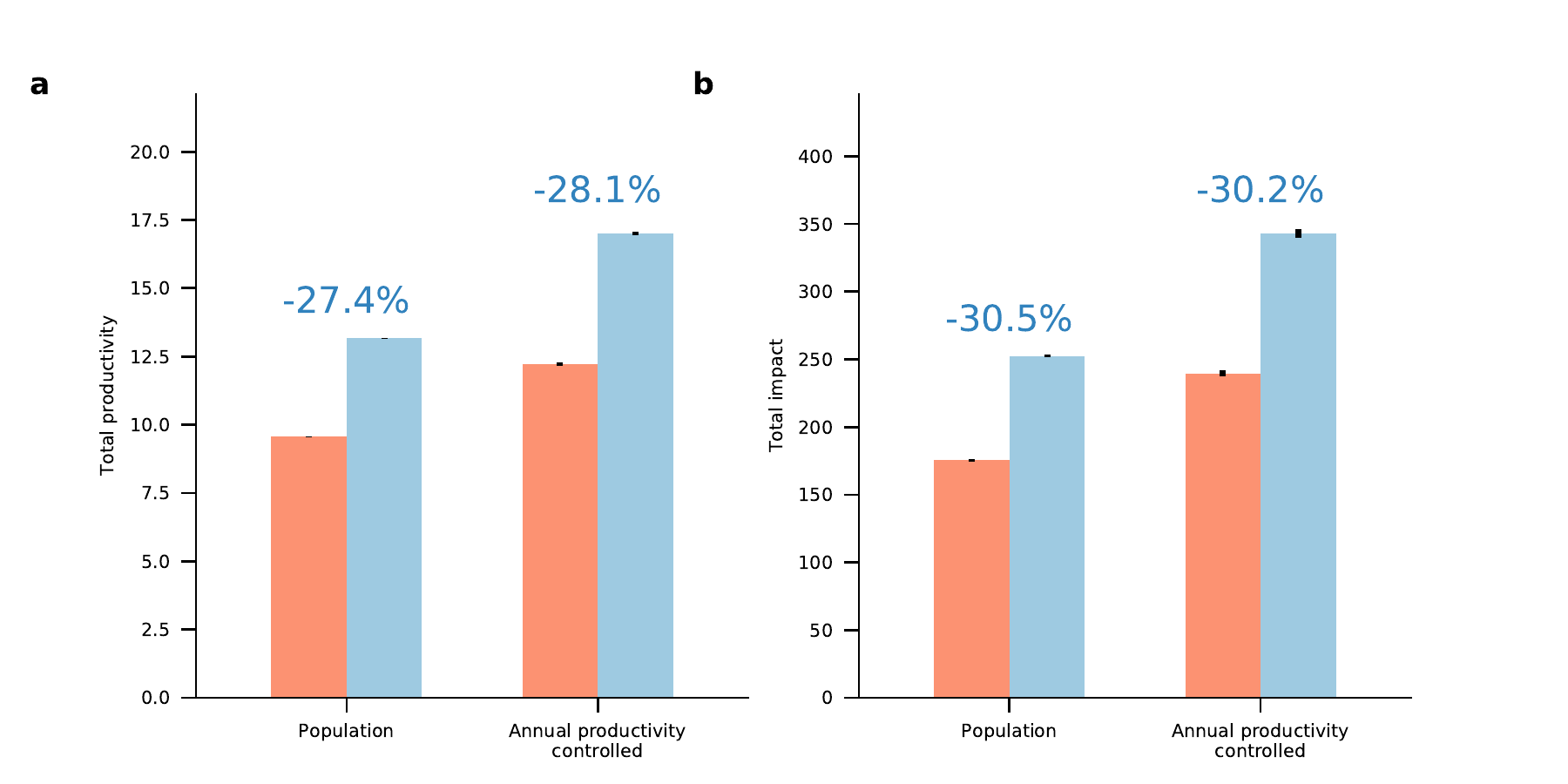}
            \caption{\textbf{Matched samples when controlling annual productivity}. Gender gaps in \textbf{a} total productivity and \textbf{b} total impact, before and after we control annual productivity between genders. The correction does not reduce gender gaps in performance.}
            \label{fig:si:control-annual-prod}
         \end{center}
    \end{figure*}

        We also ran a similar experiment controlling for the annual productivity.
        Specifically, we constructed another set of matched samples in which we identified for each female, a male author from the same country and discipline, with a nearly identical annual productivity based on grouping authors into bins by annual productivity: [0.1 papers/year, 0.2 papers/year), [0.2 papers/year, 0.3 papers/year), etc.
        The approximation occurs because annual productivity is a real-valued number.
        As seen in \figref{fig:si:control-annual-prod}{a,b}, controlling for annual productivity actually increases gender gaps in both the total productivity and total impact, although the increase is small (1.6\% and 0\% respectively).
        The lack of a significant change in the total productivity gender gap further emphasizes the importance of career length as the dominating factor.

      \subsection{Total productivity matching}\label{sec:si:prodmatching}

        Our third matching experiment controlled for the total productivity and explored the resulting change in impact.
        Specifically, we constructed another set of matched samples in which we identified for each female author, a male author from the same country, discipline, and approximately the same affiliation rank.
        In this population, the gender gap in career impact was \ImpactGapControlCountryDisciplineRank in favor of male authors.
        We then created our second experimental population, as a subset of the first, in which we matched each female author to a male author from the same country, discipline, with approximately the same affiliation rank, and with exactly the same total productivity.
       With the addition of matching on total productivity, the impact gap actually flips in favor of female scientists who receives an average of  \ImpactGapControlCountryDisciplineRankPub more citations.
        We report the mean impact gap over 100 randomized trials and the standard deviation for the impact gap is nearly zero.

    %  metrics-wos2017-x-335-341-collab-number-of-collaborators

\subsection{Relationship between productivity and number of collaborators}
\label{si:sec:collab}
\begin{figure*}[!p]
        \begin{center}
            \includegraphics{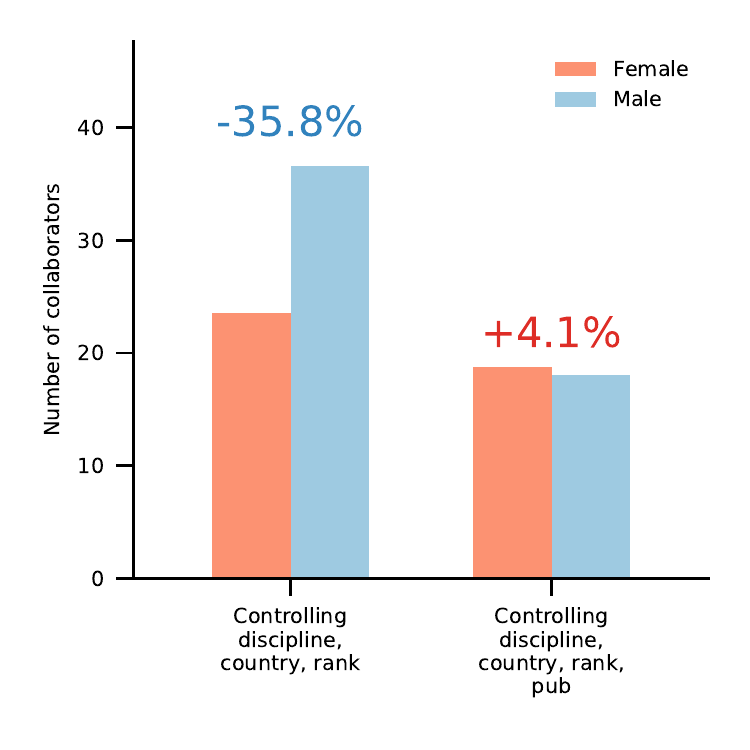}
            \caption{\textbf{Matched samples explain the average number of collaborators}. The gender gap in the number of collaborators in the matched samples when controlling for the discipline, country and affiliation rank, and when controlling for he discipline, country, affiliation rank, and number of publications.}
            \label{fig:si:collaboration}
         \end{center}
    \end{figure*}

    The gender gap in total productivity has an important implication for any reported gender gaps in collaboration and the subsequent structure of collaboration networks.
    Here, we test for this relationship by using a matching experiment in which we selected a male author from the same country, discipline, and affiliation rank.
    We then calculate the total number of collaborators that co-authored at least one publication, and find a substantial gender gap (\figref{fig:si:collaboration}{}, left): while men collaborate with an average of \CollabMaleControlCountryDisciplineRank co-authors, female authors collaborate with an average of \CollabFemaleControlCountryDisciplineRank co-authors, a gender gap of  \CollabGapControlCountryDisciplineRank.
    Next, a subset of this matched population was chosen such that the male and female authors published exactly the same number of articles throughout their careers (\figref{fig:si:collaboration}{}, right).
    We see that in this final matched population, the gender gap in number of collaborators actually switches to \CollabGapControlCountryDisciplineRankPub in favor of female authors.

    \subsection{Controlling for the dropout rate}\label{sec:si:dropoutcontrol}

        We introduce an experiment that simulates an alternative scientific population in which we manipulate the dropout rate of scientists.
        While it would be difficult to retroactively identify the potential publications a scientist would have published if their career did not terminate in a given year,
        we can more easily randomly terminate the careers of scientists earlier than reality.
        Here, we use this technique to eliminate the gender gap in dropout rate, and test for the effects on the productivity and impact gender gaps.

        As shown in the main text, Fig.\ 4\textbf{A}, the age-dependent dropout rate for women is always higher than the male dropout rate.
        To correct for this gender gap, we raise the dropout rate for male scientists to match that of the female scientists.
        Specifically, for a given year, we find the difference between the male and female dropout rates, and identify how many more men would need to dropout in order to equalize the rate.
        We then randomly select male scientists who otherwise would not have left the population the following year (we do not consider the remainder of the career length when selecting scientists) and terminate their careers.
        A selected male scientist keeps all publications until this age, while his authorships on all later publications are discarded (only the authorships are removed from the data, the career termination of a selected scientist does not affect his collaborators or citations).
        To account for the inherent randomness in this procedure, the experiment has been replicated $100$ times and we report the mean gender gaps, while the standard deviation is near zero.

\section{Detailed results on Web of Science}

\subsection{Distributions of measurements}

\begin{figure}[!p]
        \includegraphics[width = 16cm]{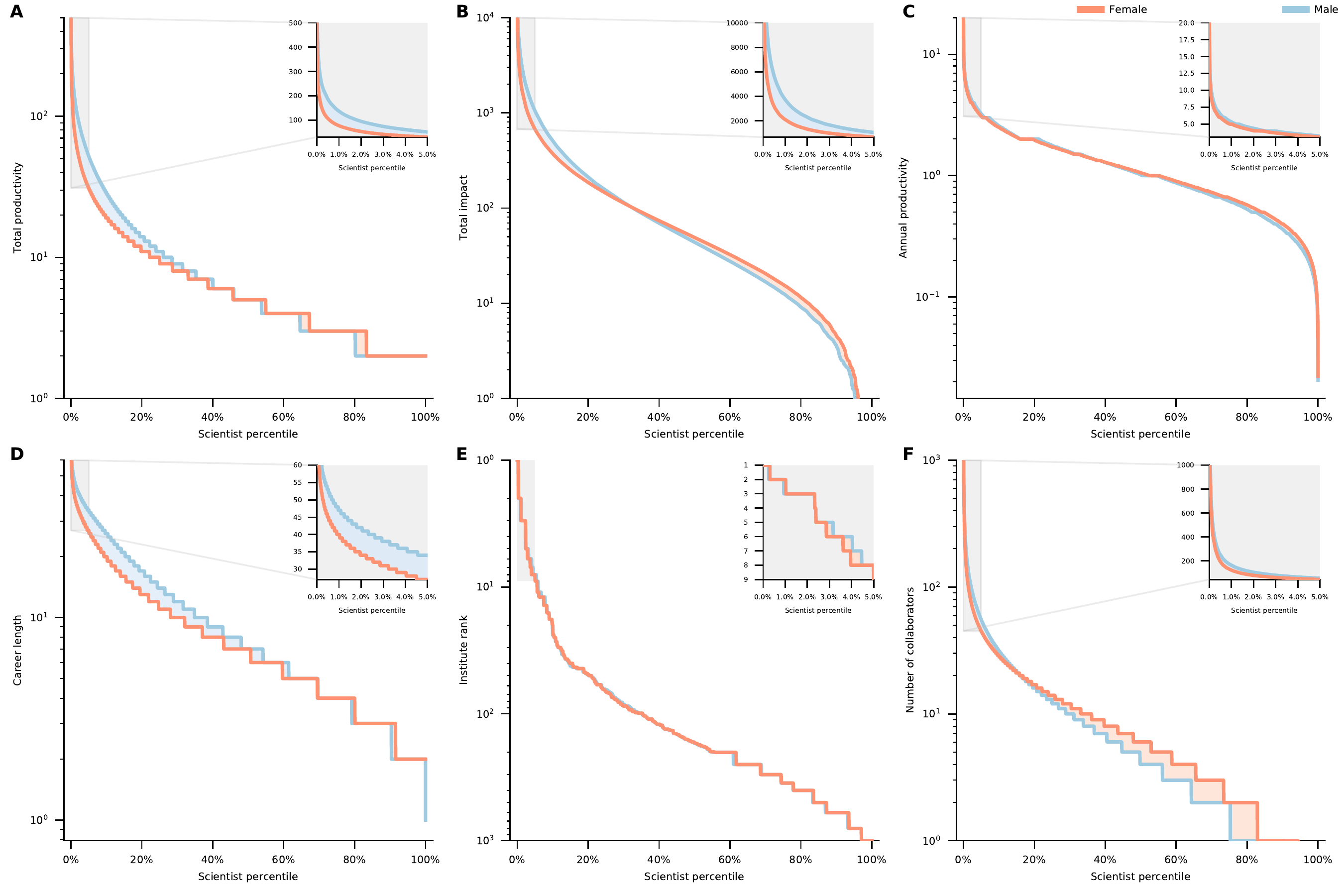}
      \caption{\textbf{Basic distributions.} Distributions of \textbf{a} total productivity, \textbf{b} total impact, \textbf{c} career length, \textbf{d} annual productivity, \textbf{e} primary institute rank, \textbf{f} number of unique collaborators.}
      \label{si:fig:metric-distributions}
    \end{figure}

    \figref{si:fig:metric-distributions}{a-d} reports the rank distributions of the four major indicators for male and female scientists.
    For each indicator type, we rank scientists from highest to lowest (denoted as the percentile of scientists with higher performance), and report the performance against percentiles.
    The difference between the rank distributions shows that, on average, male scientists have more publications and citations, and have longer careers compared to the female scientists.
    The gender inequality is most significant among top scientists (insets in all four panels).
    In contrast, male and female scientists look very similar when measured by annual productivity and citation rate.

\subsection{Statistics and gender gaps in each discipline, country, and year}

\begin{figure*}[!p]
        \begin{center}
            \includegraphics{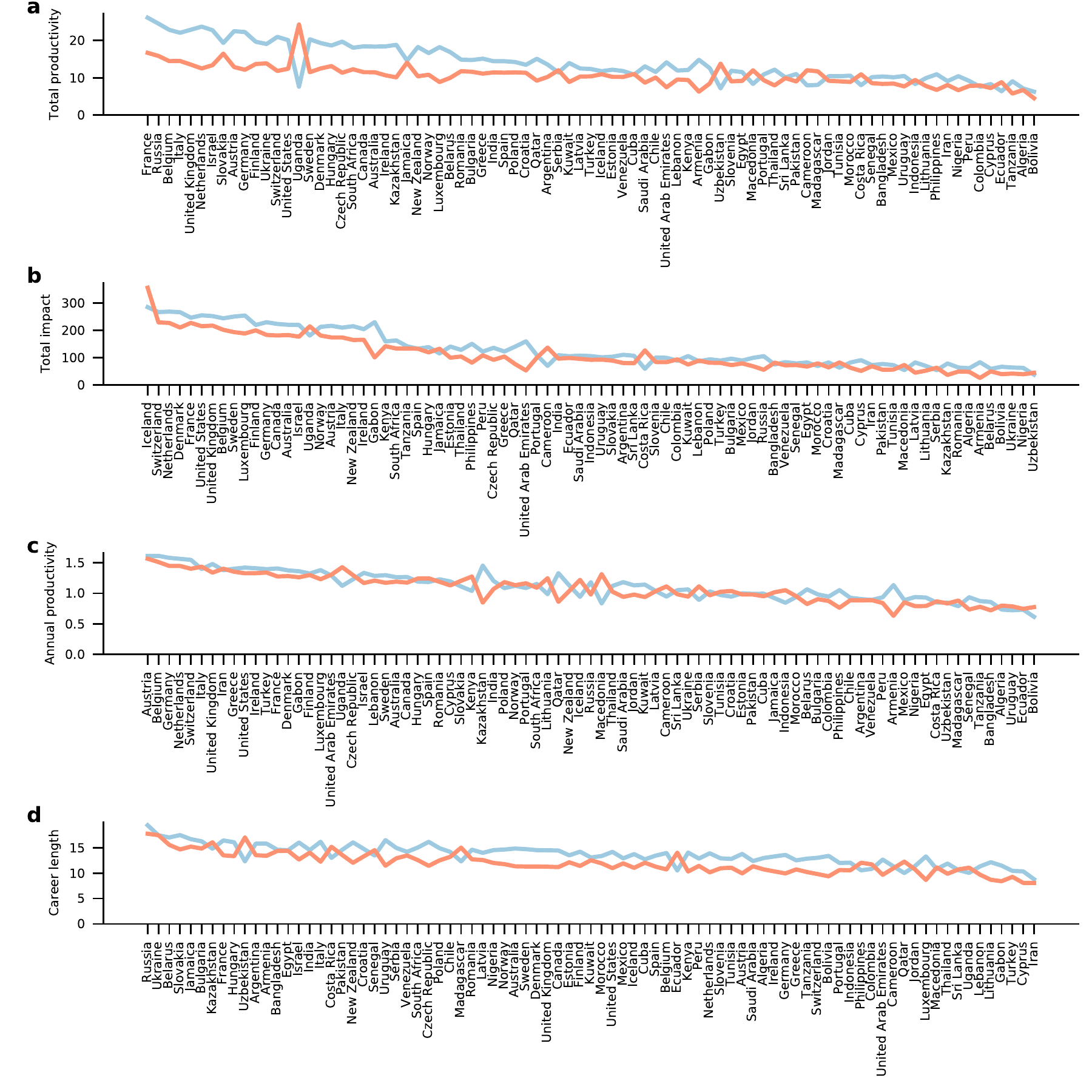}
            \caption{\textbf{The gender gap in scientific performance across countries}. The average \textbf{a} total productivity, \textbf{b} total impact, \textbf{c} annual productivity, and \textbf{d} career length among all individuals in each country.}
            \label{fig:si:country-performance}
         \end{center}
    \end{figure*}

    The gender gaps in scientific measurements across all countries (Fig.\ 2\textbf{B,G,L,Q} from the main text) is reproduced and fully labeled in \figref{fig:si:country-performance}{a-d}.
    Tables~\ref{tab:discipline-performance} and \ref{tab:country-performance} report the statistics of male and female scientists broken down by discipline and country.
    Each row reports the population size and mean performance indicators of male (in blue) and female (in orange) authors.
    The standard error is reported as one standard deviation.
    Table~\ref{tab:start-performance} and \ref{tab:end-performance} report the statistics of male and female scientists grouped by the year they start and finish their scientific careers, respectively.

\begin{figure*}[!p]
        \begin{center}
        \includegraphics{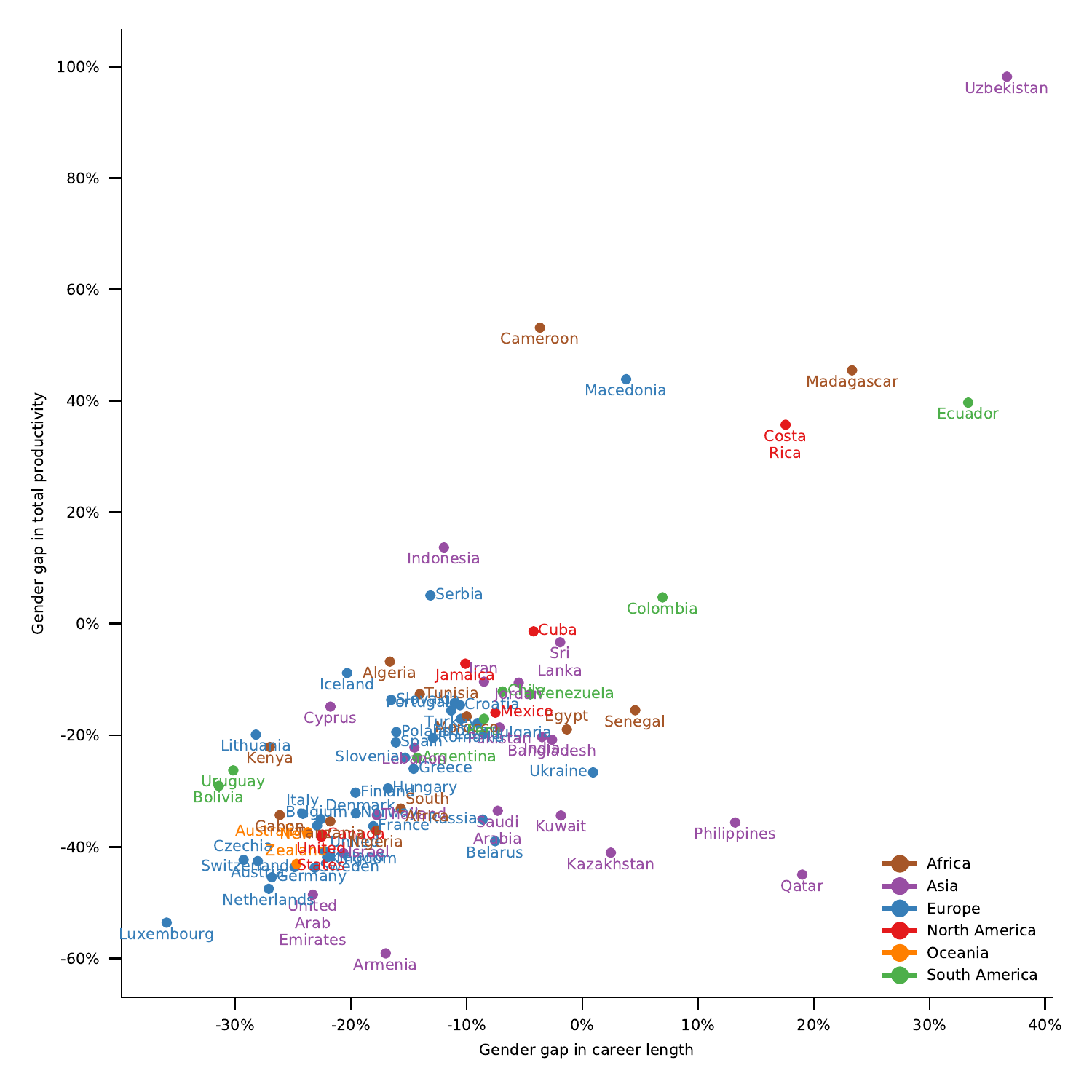}
        \caption{\textbf{The aligned gender gaps in scientific performance and career length across countries}. A full version of~\figref{fig:careerlength}{b}, demonstrating that the gender gap in career length is highly correlated with the productivity
        gap across countries.}
        \label{fig:si:aligned-gaps-in-countries}
        \end{center}
    \end{figure*}

    The detailed relationship between the gender gap in career length and total productivity across all countries is shown in \figref{fig:si:aligned-gaps-in-countries}{} as a fully labeled version of Fig.\ 3\textbf{B} from the main text.

\section{Replication in other databases}\label{sec:si:relicated}

\subsection{Microsoft Academic Graph}

\begin{figure*}[!p]
        \begin{center}
        \includegraphics{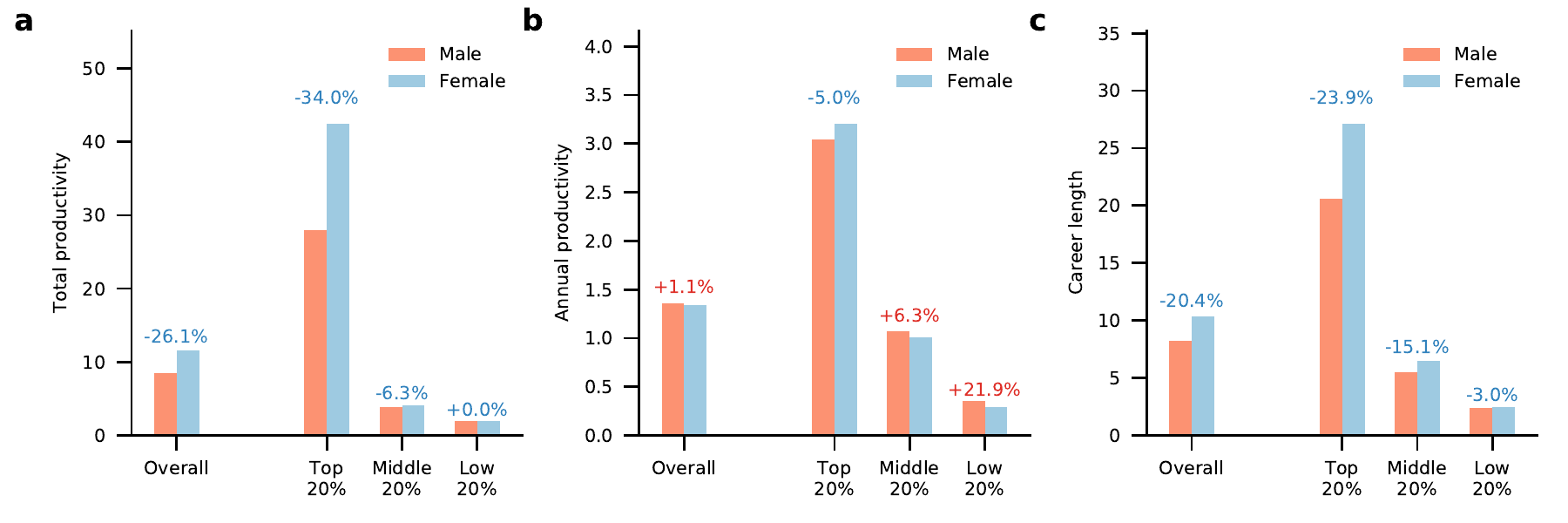}
        \caption{\textbf{The Gender Gaps in Microsoft Academic Graph}. The gender gaps in \textbf{a}, total productivity, \textbf{b}, annual productivity, and \textbf{c}, career length.
        All three gaps mirror the results for the WoS reported in the main text.}
        \label{fig:si:mag}
        \end{center}
    \end{figure*}

\begin{figure*}[!p]
        \begin{center}
        \includegraphics{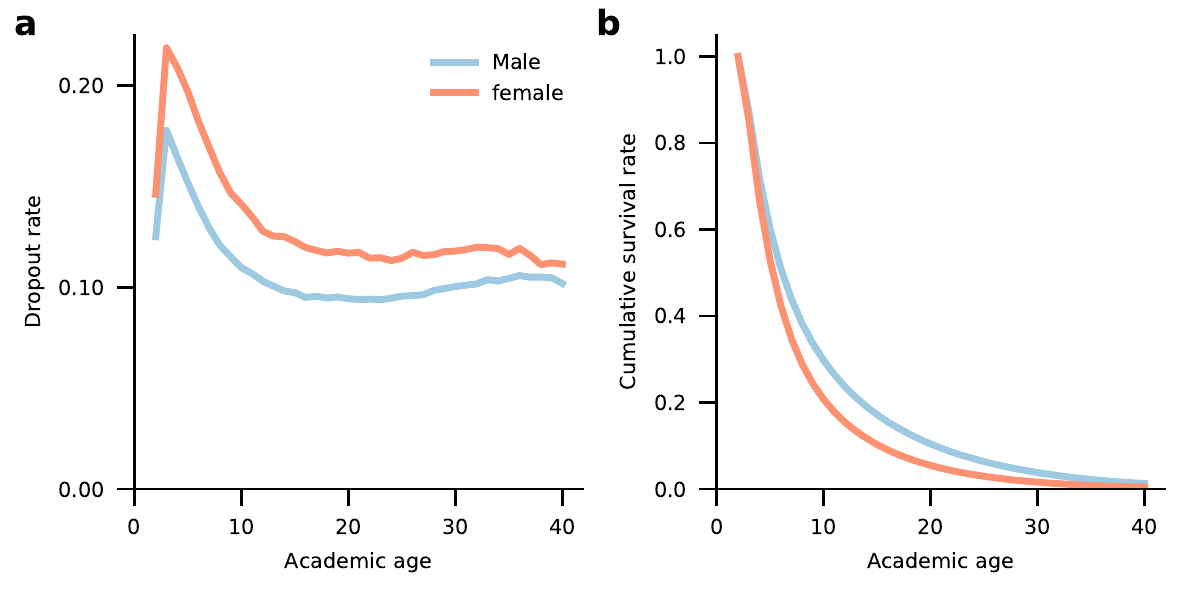}
        \caption{\textbf{Dropout and survival rates in Microsoft Academic Graph}. \textbf{a}, the dropout rate of male and female scientists at each academic age.
        \textbf{b}, the cumulative survival rate of male and female scientists at each academic age.}
        \label{fig:si:mag-dropout}
        \end{center}
    \end{figure*}

    Following the procedure for the Web of Science (Section \ref{sec:si:data}), we identified the genders of 5,856,109 male and 2,622,594 female authors who published a total of 77,642,549 articles in the MAG.
    \figref{fig:si:mag}{a-c} shows the gender gaps in total productivity, annual productivity and career length in the MAG.
    Similar to the findings reported for the WoS in the main text, we find large gender gaps in total productivity and career length, while male and female scientists differ only slightly in annual productivity.
    Likewise, we find that female scientists consistently have a higher dropout rate than male scientists (\figref{fig:si:mag-dropout}{a}) which results in a separation of the survival curves (\figref{fig:si:mag-dropout}{b}).

\subsection{DBLP}

    To prepare the DBLP data, we followed the procedure for the Web of Science (Section \ref{sec:si:data}), with the following modification.
    Because affiliation information for the DBLP is largely absent, we could not leverage location information to assist in the gender assignment.
    Instead, we compiled a list of 107,675 unique Chinese first names from the Chinese Biographical Database Project (\url{https://projects.iq.harvard.edu/cbdb/home}) and 564 unique Korean first names from wikipedia (\url{https://en.wikipedia.org/wiki/List_of_Korean_given_names}) and removed any author with a matching name from the dataset.
    After cleaning, we identified the genders of 301,150 male and 69,473 female authors who published a total of 1,740,482 articles in the DBLP.

\section{Tables and Figures}

\begin{longtable}{ p{.45\textwidth} | p{.45\textwidth} }\label{tab:reorganize-discipline}
        Web of Science category&Re-organized field\\ \hline\hline
Mathematics&a.c Mathematics\\ \hline
Computer Science&a.f Theoretical computer science\\ \hline
Physics, Thermodynamics, Mechanics, Acoustics, Crystallography&b.a Physical science - Physics\\ \hline
Chemistry, Electrochemistry, Geochemistry \& Geophysics, Spectroscopy&b.b Physical science - Chemistry\\ \hline
Oceanography&b.e Physical science - Oceanography\\ \hline
Geology&b.f Physical science - Geology\\ \hline
Meteorology \& Atmospheric Sciences&b.g Physical science - Meteorology\\
\end{longtable}

    \newpage

    \begin{longtable}{p{.2\textwidth}|p{.15\textwidth}|p{.15\textwidth}|p{.15\textwidth}|p{.15\textwidth}|p{.15\textwidth}}
        % 2019-may-round-2019-06-06/SI-wos2017-101.tex
Indicator&Female mean&Male mean&Gender gap&t-test statistic&t-test p-value\\ \hline\hline
Total productivity&\color{orange}{9.56$\pm$0.03}&\color{blue}{13.16$\pm$0.03}&-27.38\%&96.20&$\textless$1E-100\\ \hline\hline
Total impact&\color{orange}{175.49$\pm$0.86}&\color{blue}{252.35$\pm$0.87}&-30.46\%&62.05&$\textless$1E-100\\ \hline\hline
Career length&\color{orange}{9.26$\pm$0.01}&\color{blue}{11.02$\pm$0.01}&-15.91\%&109.07&$\textless$1E-100\\ \hline\hline
Annual productivity&\color{orange}{1.32$\pm$0.00}&\color{blue}{1.33$\pm$0.00}&-0.88\%&6.24&4.39E-10\\ \hline\hline

        \caption{\textbf{Academic performance.} In each row we report the average measurements ($\pm$ 1 std) of all female (orange) and male (blue) scientists for the total productivity, the total impact, career length and annual productivity. We also supply the test statistics for the difference of means between male and female scientists using the two-tailed Welch's t-test.}
        \label{tab:overall-performance}
    \end{longtable}

    \newpage

    \begin{longtable}{p{.16\textwidth}|p{.1\textwidth}|p{.2\textwidth}|p{.17\textwidth}|p{.17\textwidth}|p{.2\textwidth}}
        % 2019-may-round-2019-06-06/SI-wos2017-174-discipline.tex
Discipline&Population&Total productivity&Total impact&Career length&Annual productivity\\ \hline\hline
\multirow{3}{*}{Agronomy}&\color{blue}{26,550}&\color{blue}{13.16$\pm$0.14}&\color{blue}{117.04$\pm$1.55}&\color{blue}{12.13$\pm$0.06}&\color{blue}{1.22$\pm$0.01}\\
&\color{orange}{9,403}&\color{orange}{9.01$\pm$0.13}&\color{orange}{89.54$\pm$1.64}&\color{orange}{9.63$\pm$0.08}&\color{orange}{1.24$\pm$0.01}\\
&&-31.5\% (3E-88)&-23.5\% (1E-33)&-20.7\% (3E-122)&2.1\% (2E-02)\\
\hline
\multirow{3}{*}{Applied physics}&\color{blue}{15,662}&\color{blue}{8.74$\pm$0.12}&\color{blue}{72.59$\pm$1.30}&\color{blue}{9.05$\pm$0.06}&\color{blue}{1.31$\pm$0.01}\\
&\color{orange}{2,700}&\color{orange}{8.06$\pm$0.22}&\color{orange}{60.59$\pm$2.80}&\color{orange}{8.81$\pm$0.14}&\color{orange}{1.28$\pm$0.02}\\
&&-7.8\% (9E-03)&-16.5\% (3E-05)&-2.6\% (2E-01)&-1.9\% (2E-01)\\
\hline
\multirow{3}{*}{Biology}&\color{blue}{107,219}&\color{blue}{16.56$\pm$0.10}&\color{blue}{246.13$\pm$1.14}&\color{blue}{12.31$\pm$0.03}&\color{blue}{1.38$\pm$0.00}\\
\end{longtable}

    \newpage

    \begin{longtable}{p{.16\textwidth}|p{.1\textwidth}|p{.2\textwidth}|p{.17\textwidth}|p{.17\textwidth}|p{.2\textwidth}}
        % 2019-may-round-2019-06-06/SI-wos2017-174-country.tex
Country&Population&Total productivity&Total impact&Career length&Annual productivity\\ \hline\hline
\multirow{3}{*}{Algeria}&\color{blue}{134}&\color{blue}{7.16$\pm$0.62}&\color{blue}{60.44$\pm$8.45}&\color{blue}{12.96$\pm$0.69}&\color{blue}{0.73$\pm$0.04}\\
&\color{orange}{42}&\color{orange}{6.76$\pm$0.68}&\color{orange}{47.56$\pm$9.40}&\color{orange}{10.71$\pm$0.74}&\color{orange}{0.80$\pm$0.07}\\
&&-5.5\% (7E-01)&-21.3\% (3E-01)&-17.3\% (3E-02)&8.8\% (4E-01)\\
\hline
\multirow{3}{*}{Argentina}&\color{blue}{1,025}&\color{blue}{13.57$\pm$0.59}&\color{blue}{109.98$\pm$7.41}&\color{blue}{15.78$\pm$0.33}&\color{blue}{0.91$\pm$0.02}\\
&\color{orange}{961}&\color{orange}{10.25$\pm$0.34}&\color{orange}{79.78$\pm$3.76}&\color{orange}{13.52$\pm$0.25}&\color{orange}{0.88$\pm$0.02}\\
&&-24.4\% (2E-06)&-27.5\% (1E-04)&-14.3\% (4E-08)&-2.5\% (5E-01)\\
\hline
\multirow{3}{*}{Armenia}&\color{blue}{38}&\color{blue}{14.84$\pm$2.50}&\color{blue}{82.74$\pm$35.00}&\color{blue}{15.79$\pm$1.74}&\color{blue}{1.13$\pm$0.20}\\
\end{longtable}

    \newpage

    \begin{longtable}{p{.2\textwidth}|p{.1\textwidth}|p{.2\textwidth}|p{.15\textwidth}|p{.15\textwidth}|p{.2\textwidth}}
        % 2019-may-round-2019-06-09/SI-wos2017-174-start.tex
Year of career start&Population&Total productivity&Total impact&Career length&Annual productivity\\ \hline\hline
\multirow{3}{*}{1950-1959}&\color{blue}{47,847}&\color{blue}{29.78$\pm$0.25}&\color{blue}{619.76$\pm$5.92}&\color{blue}{20.26$\pm$0.08}&\color{blue}{1.36$\pm$0.01}\\
&\color{orange}{7,445}&\color{orange}{21.77$\pm$0.44}&\color{orange}{452.61$\pm$11.31}&\color{orange}{18.77$\pm$0.20}&\color{orange}{1.24$\pm$0.01}\\
&&-26.9\% (3E-50)&-27.0\% (3E-37)&-7.3\% (6E-13)&-8.7\% (1E-16)\\
\hline
\multirow{3}{*}{1960-1969}&\color{blue}{116,328}&\color{blue}{23.61$\pm$0.15}&\color{blue}{445.04$\pm$3.02}&\color{blue}{16.32$\pm$0.04}&\color{blue}{1.44$\pm$0.00}\\
&\color{orange}{19,439}&\color{orange}{18.94$\pm$0.27}&\color{orange}{348.78$\pm$5.22}&\color{orange}{15.75$\pm$0.10}&\color{orange}{1.34$\pm$0.01}\\
&&-19.7\% (2E-60)&-21.6\% (2E-48)&-3.5\% (1E-07)&-7.0\% (1E-29)\\
\hline
\multirow{3}{*}{1970-1979}&\color{blue}{194,606}&\color{blue}{17.59$\pm$0.07}&\color{blue}{317.77$\pm$1.87}&\color{blue}{13.75$\pm$0.02}&\color{blue}{1.39$\pm$0.00}\\
\end{longtable}

    \newpage

    \begin{longtable}{p{.2\textwidth}|p{.1\textwidth}|p{.2\textwidth}|p{.15\textwidth}|p{.15\textwidth}|p{.2\textwidth}}
        % 2019-may-round-2019-06-09/SI-wos2017-174-end.tex
Year of career end&Population&Total productivity&Total impact&Career length&Annual productivity\\ \hline\hline
\multirow{3}{*}{1950-1959}&\color{blue}{12,788}&\color{blue}{5.72$\pm$0.10}&\color{blue}{141.67$\pm$3.91}&\color{blue}{6.94$\pm$0.07}&\color{blue}{1.42$\pm$0.01}\\
&\color{orange}{2,203}&\color{orange}{5.15$\pm$0.15}&\color{orange}{131.86$\pm$7.43}&\color{orange}{6.39$\pm$0.13}&\color{orange}{1.47$\pm$0.02}\\
&&-10.0\% (2E-03)&-6.9\% (2E-01)&-7.9\% (4E-04)&3.6\% (3E-02)\\
\hline
\multirow{3}{*}{1960-1969}&\color{blue}{51,474}&\color{blue}{6.13$\pm$0.05}&\color{blue}{136.31$\pm$1.97}&\color{blue}{6.86$\pm$0.03}&\color{blue}{1.45$\pm$0.00}\\
&\color{orange}{8,874}&\color{orange}{5.19$\pm$0.09}&\color{orange}{115.59$\pm$3.60}&\color{orange}{6.23$\pm$0.07}&\color{orange}{1.47$\pm$0.01}\\
&&-15.3\% (9E-24)&-15.2\% (6E-08)&-9.2\% (4E-17)&1.6\% (5E-02)\\
\hline
\multirow{3}{*}{1970-1979}&\color{blue}{100,433}&\color{blue}{6.56$\pm$0.04}&\color{blue}{135.54$\pm$1.42}&\color{blue}{7.03$\pm$0.02}&\color{blue}{1.42$\pm$0.00}\\
\end{longtable}

    \newpage

    \begin{longtable}{p{.2\textwidth}|p{.1\textwidth}|p{.2\textwidth}|p{.15\textwidth}|p{.15\textwidth}|p{.2\textwidth}}
        % 2019-may-round-2019-06-09/SI-wos2017-174-rank.tex
Institute rank&Population&Total productivity&Total impact&Career length&Annual productivity\\ \hline\hline
\multirow{3}{*}{1-19}&\color{blue}{545}&\color{blue}{29.57$\pm$0.39}&\color{blue}{756.34$\pm$9.86}&\color{blue}{15.25$\pm$0.08}&\color{blue}{1.76$\pm$0.01}\\
&\color{orange}{221}&\color{orange}{17.92$\pm$0.40}&\color{orange}{500.26$\pm$12.69}&\color{orange}{12.02$\pm$0.11}&\color{orange}{1.55$\pm$0.01}\\
&&-39.4\% (2E-117)&-33.9\% (9E-63)&-21.2\% (3E-110)&-11.8\% (2E-30)\\
\hline
\multirow{3}{*}{20-48}&\color{blue}{280}&\color{blue}{27.09$\pm$0.29}&\color{blue}{544.44$\pm$7.80}&\color{blue}{15.02$\pm$0.09}&\color{blue}{1.65$\pm$0.01}\\
&\color{orange}{108}&\color{orange}{16.61$\pm$0.35}&\color{orange}{357.64$\pm$8.66}&\color{orange}{11.53$\pm$0.11}&\color{orange}{1.50$\pm$0.01}\\
&&-38.7\% (2E-107)&-34.3\% (2E-54)&-23.2\% (1E-143)&-9.2\% (1E-19)\\
\hline
\multirow{3}{*}{49-86}&\color{blue}{913}&\color{blue}{27.56$\pm$0.24}&\color{blue}{537.63$\pm$6.52}&\color{blue}{15.40$\pm$0.09}&\color{blue}{1.64$\pm$0.01}\\
\end{longtable}

    \newpage

    \begin{longtable}{p{.2\textwidth}|p{.1\textwidth}|p{.2\textwidth}|p{.15\textwidth}|p{.15\textwidth}|p{.2\textwidth}}
        % 2019-may-round-2019-06-09/SI-wos2017-174-collab.tex
Number of collaborators&Population&Total productivity&Total impact&Career length&Annual productivity\\ \hline\hline
\multirow{3}{*}{0}&\color{blue}{103,414}&\color{blue}{3.04$\pm$0.01}&\color{blue}{27.76$\pm$0.18}&\color{blue}{6.77$\pm$0.02}&\color{blue}{1.00$\pm$0.00}\\
&\color{orange}{23,317}&\color{orange}{2.92$\pm$0.02}&\color{orange}{26.91$\pm$0.33}&\color{orange}{6.20$\pm$0.03}&\color{orange}{1.00$\pm$0.01}\\
&&-3.8\% (1E-10)&-3.1\% (1E-02)&-8.3\% (2E-38)&-0.2\% (8E-01)\\
\hline
\multirow{3}{*}{1}&\color{blue}{171,648}&\color{blue}{3.26$\pm$0.00}&\color{blue}{31.07$\pm$0.16}&\color{blue}{5.96$\pm$0.01}&\color{blue}{1.13$\pm$0.00}\\
&\color{orange}{47,362}&\color{orange}{3.20$\pm$0.01}&\color{orange}{30.91$\pm$0.29}&\color{orange}{5.76$\pm$0.02}&\color{orange}{1.11$\pm$0.00}\\
&&-2.1\% (2E-09)&-0.5\% (6E-01)&-3.3\% (1E-12)&-1.9\% (3E-07)\\
\hline
\multirow{3}{*}{2}&\color{blue}{122,155}&\color{blue}{4.03$\pm$0.01}&\color{blue}{43.54$\pm$0.27}&\color{blue}{6.91$\pm$0.02}&\color{blue}{1.13$\pm$0.00}\\
\end{longtable}

    \newpage

    \begin{figure*}[!p]
    \begin{center}
        \includegraphics[width=19cm]{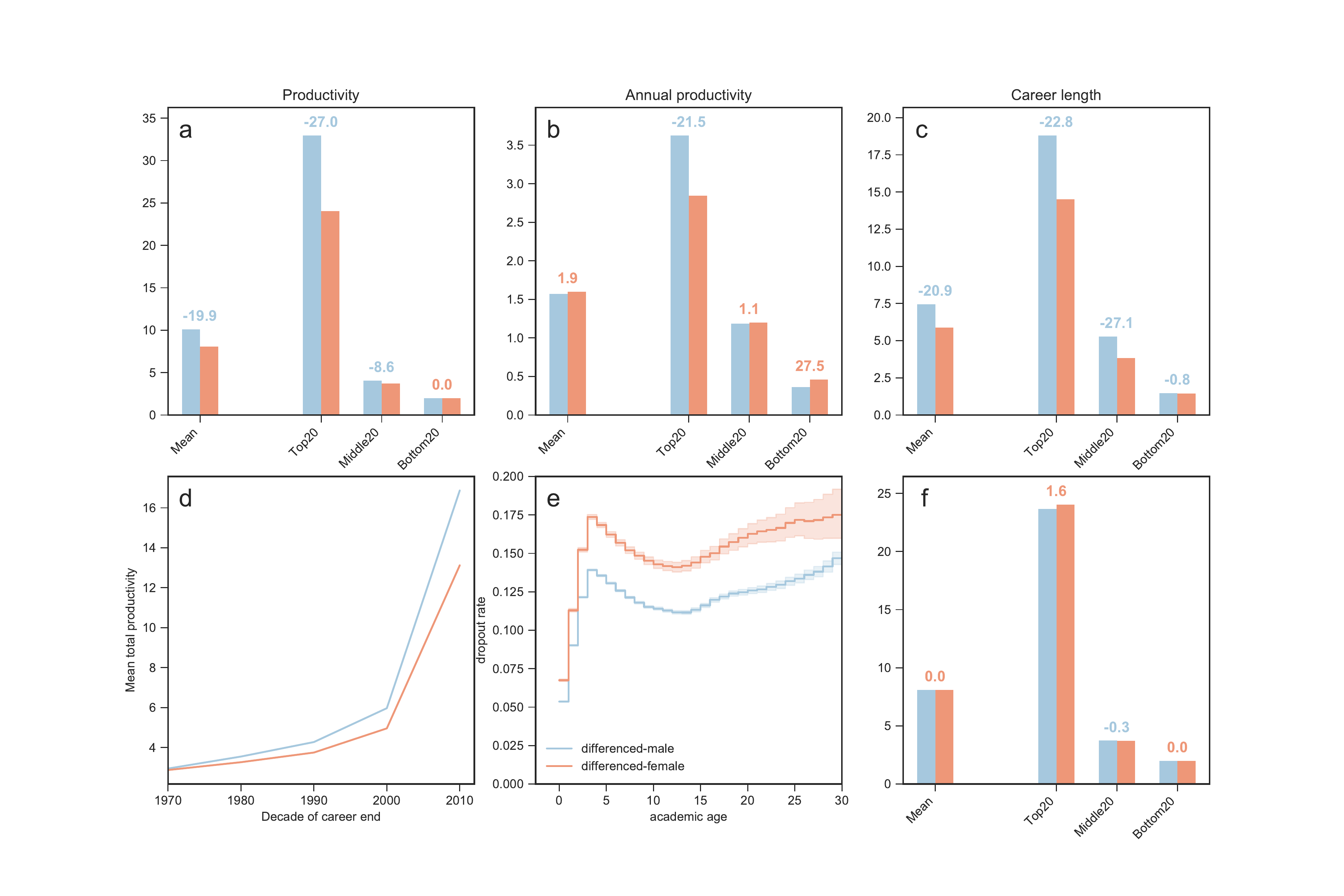}
    	\caption{\textbf{The Gender Gaps in DBLP}. \textbf{a}, The productivity puzzle as demonstrated by the difference in total productivity of an author during his/her career.
        \textbf{b}, the annual productivity is nearly identical for male and female authors.
        \textbf{c}, the difference in career length for male and female authors.
        \textbf{d}, the gender gap in productivity is growing over that last 40 years.
        \textbf{e}, female authors have higher dropout rate than male authors at all stages of their careers.
        \textbf{f}, a matching experiment eliminates the productivity gap.
         All conclusions qualitatively mirror the results for the WoS reported in the main text.
        }
     \label{fig:si:dblp}
     \end{center}
    \end{figure*}

\end{document}